\documentclass[reprint, longbibliography,
 amsmath,amssymb,
 aps, prx,
]{revtex4-2}

\usepackage{amsmath,amssymb,amsfonts}

\usepackage{graphicx}
\usepackage[caption=false]{subfig}
\usepackage{booktabs,array,makecell}

\usepackage{algorithm2e}
\usepackage[version=4]{mhchem}
\usepackage{siunitx}

\usepackage{physics}
\usepackage{quantikz}

\DeclareUnicodeCharacter{2212}{\ensuremath{-}}
\usepackage{microtype}
\usepackage{xcolor}
\usepackage{hyperref}
\hypersetup{
    colorlinks=true,
    linkcolor=blue,
    citecolor=blue,
    urlcolor=blue
}

\def\quantinuumCambridge{Quantinuum, Terrington House, 13-15 Hills Road, Cambridge CB2 1NL, UK}

\begin{document}

\preprint{APS/123-QED}

\title{\textbf{Split-Evolution Quantum Phase Estimation for Particle-Conserving Hamiltonians}}% 

\author{Megan Cerys Rowe}
\author{Carlo A. Gaggioli}
\author{Ludmila Szulakowska}
\author{David Muñoz Ramo}
\author{David Zsolt Manrique}
\email{david.zsolt.manrique@quantinuum.com}
\affiliation{\quantinuumCambridge}

\date{\today} % Leave empty to omit a date

\begin{abstract}
We present a hardware demonstration and resource analysis of split-evolution quantum phase estimation (SE-QPE) on a Quantinuum System Model H2 quantum computer.
SE-QPE is a modification to canonical QPE for particle-conserving Hamiltonians in which controlled time evolution is replaced by CSWAP-based interference
between a target register and a reference register. For factorizations of time evolution with a shared eigenbasis,
SE-QPE preserves the phase-register outcome distribution of canonical QPE and, unlike with compute--uncompute substitutions,
it remains compatible with non-exact eigenstates. The substitution removes controlled-simulation
overhead and enables parallel evolution on two registers, reducing the depth of each phase-kickback block. Resource analysis
for Trotterized double-factorized chemistry Hamiltonians shows that the substitution becomes increasingly favorable at
higher phase powers and combining QPE and SE-QPE implementations can be a useful option. Over a range of FeMoco active spaces, SE-QPE reduces time evolution resources, with asymptotic reductions of about 33\% in CX count,
25\% in $T$ count, and an asymptotic depth ratio of $3/N$ for CX layers.
On Quantinuum H2-2, a four-qubit model ethylene demonstration with explicit inverse QFT and repeated phase-kickback steps up to 8 phase bits
yields distinct energies and shows the auxiliary registers provide useful error detection filters.\end{abstract}

\maketitle

\section{Introduction}
\label{sec:intro}
Accurate ground- and excited-state energies are a fundamental requirement for predictive modeling of many physical and chemical processes. A practically feasible method for high-accuracy energies would improve, for example, predictions of formation energies and thermodynamic stability \cite{Hautier_2012, Jain_2013}, adsorption energies and activation barriers determining kinetic rates in catalysis and reaction networks \cite{Norskov_2009,ChenXuMavrikakis_2021}, as well as excitation energies and response properties relevant to optical spectra and photochemistry \cite{DreuwHeadGordon_2005}.
Quantum phase estimation (QPE) is a core algorithm for extracting spectral information from a unitary and underlies many proposals for quantum advantage in molecular electronic structure \cite{AspuruGuzik_2005,Reiher_2017} and other domains \cite{Kitaev_1995,Cleve_1998}. Given access to the time evolution operator $U(\tau)=e^{-i\tau H}$ for evolution time $\tau$ of a Hamiltonian $H$, QPE estimates eigenphases $\Phi_k(\tau)$ defined by $U(\tau)\ket{\chi_k}=e^{2\pi i \Phi_k(\tau)}\ket{\chi_k}$, where $\ket{\chi_k}$ are eigenstates and $\Phi_k(\tau)= -\frac{\tau E_k}{2\pi}\ \bmod 1$.

In practice, canonical QPE with an $M$-qubit phase register requires controlled applications of the powers $U(\tau)^{2^j}$ for $j=0,\ldots,M-1$. When these powers are realized by repeating a base step $U(\tau)$, the total number of base evolutions is $K=\sum_{j=0}^{M-1}2^j=2^M-1$. For chemistry Hamiltonians with many terms and nontrivial circuit constructions for $e^{-i\tau H}$ \cite{Reiher_2017,Motta_2021}, controlled time evolution is a major contributor to the total resources and circuit depth of QPE. These constraints have limited experimental demonstrations of QPE-based energy estimation to small systems and simplified models \cite{Lanyon_2010} and we expect it to remain a major difficulty in the near future.

Many avenues have been explored to reduce QPE resource requirements. One direction targets more efficient circuit implementations of $U(\tau)$ \cite{Motta_2021,Lee2021,vonBurg2021}. Another direction modifies the phase-estimation routine itself, for example, by reducing the number of ancillas and removing the inverse QFT in favor of repeated sampling and classical post-processing \cite{GriffithsNiu_1996,Dobsicek_2007,WiebeGranade_2016}. Iterative phase estimation (IQPE) and Bayesian variants avoid an explicit inverse QFT, and allow implementations with a single time evolution step per circuit, but still rely on controlled applications of $U(\tau)^{2^j}$ in their standard forms \cite{Dobsicek_2007,WiebeGranade_2016}. Related approaches extract eigenvalues from time-series data such as overlaps $\bra{\psi}U(\tau)\ket{\psi}$ using classical signal processing \cite{Somma_2019,LinTong_2022,DingLin_2023}. While the complex overlap is most directly accessed by Hadamard-test-type circuits, there are constructions that move conditionality into state preparation and unpreparation to avoid controlled time evolution \cite{LinTong_2022}. Even phase-less signals of $|\bra{\psi}U(\tau)\ket{\psi}|$ hold spectral information, which can be obtained with control-free protocols, and can be combined with phase-retrieval post-processing to recover spectral features, however with increased sampling costs \cite{Clinton_2024,phaselift}. The Bayesian phase-difference estimation (BPDE) approach instead shifts conditionality to excitation operators, enabling direct estimation of energy gaps via control-free time evolution \cite{Sugisaki_2021,Sugisaki_2023}. Complementary efforts target control overhead directly, including hardware-efficient constructions that realize effective controlled evolutions using more local primitives \cite{Schiffer_2025}. Separately from QPE, interferometric primitives based on CSWAP gates have been explored as a means to substitute controlled unitaries with uncontrolled ones \cite{Ekert_2002,Tazi_2024}.
These primitives have been explored in the context of a non-orthogonal quantum eigensolver with implementations limited to unitaries that act trivially on the vacuum state \cite{Huggins_NO_2020, Baek_2023}. Parallelization of QPE has also been proposed \cite{braun2022errorresilientquantumamplitude}, but schemes that rely on exact eigenstates can fail when the input is a superposition, which is the typical practical case.

The prior molecular phase-estimation literature on hardware remains sparse and methodologically diverse.
O'Malley \emph{et al.} reported a scalable molecular implementation without exponentially costly precompilation,
but for a two-qubit $\mathrm{H}_2$ Hamiltonian and using a modification of Kitaev's iterative phase estimation algorithm rather
than an explicit inverse QFT \cite{OMalley_2016}. The same work also notes that earlier chemistry phase-estimation
demonstrations used non-scalable configuration-basis encodings in which the propagator was classically exponentiated
before the experiment \cite{OMalley_2016}. More recently, a multi-ancilla molecular QPE was reported in a two-qubit system register setting \cite{Tranter_2025},
while other hardware studies have instead focused on Bayesian or statistical
phase-estimation variants rather than canonical QPE with an explicit inverse QFT \cite{Paesani_2017,Yamamoto_2024,Blunt_2023}.
From our literature investigation the present experiment is therefore the first demonstration of chemistry-based phase estimation on a
four-qubit system register with an explicit inverse QFT and without exponentially expensive precompilation.

In this work we present \emph{split-evolution quantum phase estimation} (SE-QPE),
a phase-estimation method that preserves the phase-register measurement statistics of canonical
QPE under explicit conditions while eliminating controlled time evolution and reducing circuit depth through two-register parallelization.
The central ingredient is a CSWAP control gadget (Fig.~\ref{fig:cswap-gadget}), hereafter the CSWAP-gadget,
which replaces the controlled-$U$ block in QPE (see circuit implementation in Fig.~\ref{fig:qppe_circuit}) and acts between a target register and a reference register.
For molecular electronic Hamiltonians, the reference can be chosen as the vacuum, the phase of a unitary acting on this state is classically computable
and can be compensated by a single-qubit rotation, addressing the absence of a direct phase reference in control-free settings \cite{Tang_2025}.
The reference register also enables practical error detection at each gadget application. 

The remainder of the paper is organized as follows.
Section~\ref{sec:pcpe} describes the CSWAP-gadget and its embedding into QPE.
Section~\ref{sec:resources} and Section~\ref{sec:femoco_resources} analyze resource requirements under Trotterized implementations of $e^{-i\tau H}$ using a double-factorized representation of $H$, with particular focus on the example of FeMoco \cite{Motta_2021}.
Section~\ref{sec:experiments} reports simulations and hardware experiments for a model Hamiltonian of C$_2$H$_4$, and Section~\ref{sec:discussion} concludes with outlook and extensions.

\begin{figure}[h]
\begin{quantikz}
&\ctrl{1}                &\qw \\
&\gate[2]{{S}_U} &\qw \\
&\qw                     &\qw 
\end{quantikz}
=
\begin{quantikz}
&\ctrl{1}  & \gate{P(\theta)}      &\ctrl{1} &\qw\\
&\swap{1}  & \gate{U_A^{\dagger}}  &\swap{1} &\qw\\
&\targX{}  & \gate{U_B}            &\targX{} &\qw
\end{quantikz}
\caption{The CSWAP-gadget, a circuit-level construction of ${S}_U$ for a factorization $U=U_AU_B$. The $N$ controlled-SWAPs implement a conditional exchange of the two $N$-qubit registers. The single-qubit gate $P(\theta)$ implements a phase shift $e^{2\pi i\theta}$ on the $\ket{1}$ state.}
\label{fig:cswap-gadget}
\end{figure}
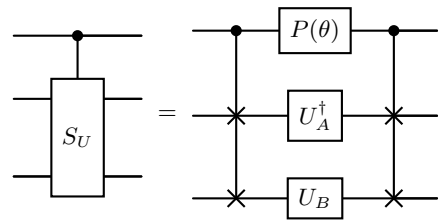

\begin{figure}[h]
\resizebox{0.48\textwidth}{!}{
\begin{quantikz}[column sep=0.35cm,row sep=0.25cm]
\lstick[wires=4]{$\ket{0}^{\otimes M}$}
  & \gate{H} & \qw & \qw
  & \ \ldots\ & \ctrl{4}\gategroup[wires=6,steps=1,style={draw,dashed,rounded corners,inner sep=-1pt}]{} 
  & \gate[wires=4]{\mathrm{QFT}^\dagger}
  & \meter{} & \cw \rstick{$m_{M-1}$} \\
\setwiretype{n}
  &
  \vdots
  &&&&&& 
  \vdots
  &  \\
\qw
  & \gate{H} & \qw
  & \ctrl{2}\gategroup[wires=4,steps=1,style={draw,dashed,rounded corners,inner sep=-1pt}]{} 
  & \ \ldots\
  && \ghost{\mathrm{QFT}^\dagger}
  & \meter{} & \cw \rstick{$m_1$} \\
\qw
  & \gate{H}
  & \ctrl{1}\gategroup[wires=3,steps=1,style={draw,dashed,rounded corners,inner sep=-1pt}]{} 
  & \qw & \ \ldots\
  && \ghost{\mathrm{QFT}^\dagger}
  & \meter{} & \cw \rstick{$m_0$} \\
\lstick{$\ket{\psi}$}
  & \qwbundle{N}
  & \gate[2]{{S}_{U^{2^0}}} 
  & \gate[2]{{S}_{U^{2^1}}} 
  &  \ \ldots\
& \gate[2]{{S}_{U^{2^{M-1}}}} 
  & \qw 
\\
\lstick{$\ket{\phi}$}
  & \qwbundle{N}
  && \qw & \ \ldots\ 
  & \qw
  & \qw 
\end{quantikz}
}

\caption{SE-QPE circuit with the CSWAP-gadget. A canonical quantum phase estimation circuit in which each controlled-$U^{2^j}$ block is replaced by a two-register CSWAP-gadget ${S}_{U^{2^j}}$ acting on a target register initialized in $\ket{\psi}$ and a reference register initialized in $\ket{\phi}$.}
\label{fig:qppe_circuit}
\end{figure}
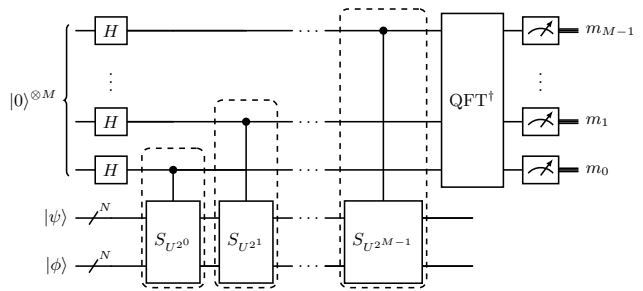

\begin{figure*}[t]
\centering
\resizebox{0.98\textwidth}{!}{$
\begin{quantikz}[wire types={n,q,q,n}, classical gap=0.1cm]
&&&& \\
\lstick{$\ket{+}_C$}    & \qw          &\gategroup[wires=2,steps=2,style={draw,dashed,rounded corners,inner sep=10pt}]{}& \ctrl{1}             & \qw & \qw  \rstick[2]{$\ket{\Psi}$}    \\
\lstick{$\ket{\psi}_A$} && \qwbundle{N} & \gate[1]{\text{$U$}} & & \qw \\
&&&& 
\end{quantikz}
\ \to \ 
\begin{quantikz}[wire types={q,q,q,q}, classical gap=0.1cm]
\setwiretype{n}
&&\gategroup[wires=4,steps=9,style={draw,dashed,rounded corners,inner sep=10pt}]{}& 
\lstick{$\ket{\overline{0}}_F$}
& \setwiretype{b}
& \gate{\overline{X}}
& \ctrl[style={dotted}]{1}
& \qw       
& \ctrl[style={dotted}]{1}
& \gate{\overline{X}} 
& \meterD{00...0} & \setwiretype{n}
\\
\lstick{$\ket{+}_C$}              
&&& \qw          
& \qw
& \ctrl{-1}
& \ctrl{1} 
& \gate{P(\theta)}       
& \ctrl{1} 
& \ctrl{-1}
& \qw
&& \rstick[2]{$I_C \otimes \Gamma_A\,\ket{\Psi}$} 
\\
\lstick{$\ket{\psi}_A$}    
&&& \qw
& \qw
& \qwbundle{N}  
& \swap{1} 
& \gate{U_A^{\dagger}}           
& \swap{1} 
& \qw 
& \qw 
&& \qw 
\\
\setwiretype{n}
&&&
\lstick{$\ket{\overline{0}}_B$} 
&
 \setwiretype{q}
& \qwbundle{N}
& \targX{} 
& \gate{U_B} 
& \targX{} 
& \meterD{00...0} &\setwiretype{n}&
\end{quantikz}
$}
\caption{Vacuum-reference substitution for a controlled-$U$ block. The reference register $B$ is initialized in the vacuum state $\ket{\overline 0}$ and can be measured to detect leakage out of the reference subspace. The optional register $F$ prepares a cat (GHZ) state to parallelize the controlled-SWAP layer. Measuring $F$ and post-selecting the all-zero outcome provides an additional error detecting check. The unitary $\Gamma_A$ denotes a diagonal operator in the eigenbasis of $U$ that accumulates phases on the system registers but does not affect the phase-register measurement statistics (see Eq.~\eqref{eq:SU_product_action}).}
\label{fig:qppe_vacuum_circuit}
\end{figure*}
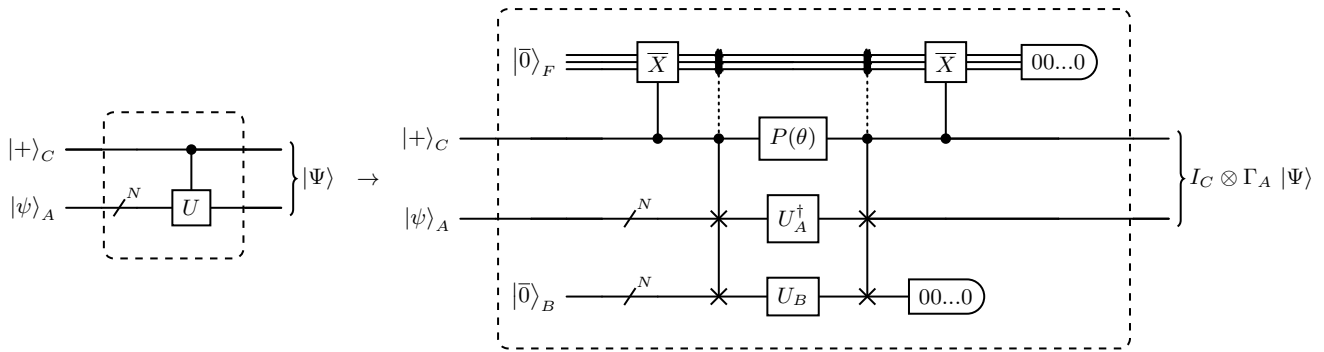

\section{Circuit substitutions for QPE}
\label{sec:pcpe}
We describe the CSWAP-gadget ${S}_U$, which employs a cascade of CSWAP gates; two system register lanes (a state lane and a reference lane) and a reference correction phase gate 
to replace a controlled-$U$ block in phase-estimation circuits for particle-conserving Hamiltonians.
Suppose a unitary $U$, whose eigenphases we wish to estimate, is factorized as
\begin{equation}
U = U_A U_B,
\end{equation}
where $U_A$ and $U_B$ are unitaries implementable with circuits. The CSWAP-gadget in Fig.~\ref{fig:cswap-gadget} defines a circuit block $S_U$ that implements the unitary 
\begin{equation}
S_U
=
\ket{0}\bra{0}\otimes \left(U_A^\dagger\otimes U_B\right)
+
e^{2\pi i\theta}\ket{1}\bra{1}\otimes \left(U_B\otimes U_A^\dagger\right),
\label{eq:SU_block}
\end{equation}
where the first tensor factor is the control qubit, the last two factors correspond to the two $N$-qubit system registers, and $\theta$ is an adjustable phase implemented by $P(\theta)$ in Fig.~\ref{fig:cswap-gadget}. The two control branches differ only by exchanging which register lane receives $U_A^\dagger$ and which receives $U_B$. The output of $S_U$ is not, in general, identical to a directly controlled unitary because both conditional branches apply different rotations to the system registers. A key sufficient condition under which $S_U$ can provide the phase-kickback that is necessary for QPE is that $U_A$ and $U_B$ share an eigenbasis $\{\ket{\chi_k}\}$, i.e.,
\begin{equation}
U_A\ket{\chi_k}=e^{2\pi i\alpha_k}\ket{\chi_k},
\qquad
U_B\ket{\chi_k}=e^{2\pi i\beta_k}\ket{\chi_k},
\label{eq:UAUB_share_eigs}
\end{equation}
so the eigenphase of $U$ on $\ket{\chi_k}$ is $\Phi_k=\alpha_k+\beta_k$. This condition is satisfied, for example, when $U_A$ and $U_B$ are commuting functions of the same Hamiltonian.

Acting on an eigenstate pair $\ket{+}\ket{\chi_k}\ket{\chi_l}$ we obtain
\begin{equation}
\begin{aligned}
S_U\ket{+}&\ket{\chi_k}\ket{\chi_l}
=\\&
=e^{2\pi i(\beta_l-\alpha_k)}
\frac{1}{\sqrt{2}}
\Bigl(
\ket{0}
+
e^{2\pi i(\theta+\Phi_k-\Phi_l)}\ket{1}
\Bigr)
\ket{\chi_k}\ket{\chi_l}
\end{aligned}
\label{eq:SU_general_action}
\end{equation}
where the phase difference $\Phi_k-\Phi_l$ appears as phase-kickback expressed onto the ancilla qubit. By superposition, any two-lane state expanded as $\sum_{k,l} d_{kl}\ket{\chi_k}\ket{\chi_l}$ can be used to target phase differences in QPE.
In particular, for an input $\ket{\psi}\ket{\phi}$ with
$\ket{\psi}=\sum_k c_k\ket{\chi_k}$, $\ket{\phi}=\sum_l f_l\ket{\chi_l}$ and $d_{kl}=c_k f_l$ we can find 
\begin{equation}
\begin{aligned}
S_U&\ket{+}\ket{\psi}\ket{\phi}
=\\&
=
\sum_{k,l} d_{kl}\,
e^{2\pi i(\beta_l-\alpha_k)}
\frac{\ket{0}
+
e^{2\pi i(\theta+\Phi_k-\Phi_l)}\ket{1}}{\sqrt{2}}
\ket{\chi_k}\ket{\chi_l},
\end{aligned}
\label{eq:SU_product_action}
\end{equation}
where strictly the factor $e^{2\pi i(\beta_l-\alpha_k)}$ changes only the phase of the coefficients of the output state,
\begin{equation}
d_{kl}\mapsto d'_{kl}=d_{kl}e^{2\pi i(\beta_l-\alpha_k)},
\end{equation} therefore  the probabilities of the eigenbasis components $|d'_{kl}|^2=|d_{kl}|^2$ are unchanged.
Under successive applications of the CSWAP-gadget (as in the SE-QPE), the coefficient phases accumulate only in the state and reference registers and do not appear in the phase-register. Consequently,  measurement outcome statistics, which depend on the probabilities $|d_{kl}|^2$ and the phase register carrying the phase difference $(\theta+\Phi_k-\Phi_l) \ \mathrm{mod}\ 1$, remains unaffected. Therefore by providing phase kickback onto the phase register the CSWAP-gadget can replace controlled-$U$ blocks in typical phase-estimation routines, including canonical and iterative QPE. The gadget also naturally supports phase-difference estimation between two prepared input states \cite{Sugisaki_2023}. For chemistry applications with a single target state, a convenient strategy is to use an accessible reference eigenstate. If one lane is prepared in a fixed reference eigenstate $\ket{\chi_{l'}}$ with known $\Phi_{l'}$, then the estimated phase is
$\left(\theta+\Phi_k-\Phi_{l'}\right)\ \bmod 1$,
i.e., $\Phi_k$ up to a known offset.

Beyond removing controlled time evolution, the two-register structure provides a depth advantage, as within each gadget, $U_A^\dagger$ and $U_B$ act on different registers and can be executed in parallel. If the implementation depth scales linearly with simulated time \cite{berry2007efficient} (as for Trotterized circuits), a balanced split between the registers can reduce the two-qubit depth of the evolution segment by up to a factor of two. For example, for the $j$th phase bit one often implements
\begin{equation}
U(\tau)^{2^j} = U(2^j\tau) = U(2^{j-1}\tau)\,U(2^{j-1}\tau),
\end{equation}
and can take $U_A=U_B=U(2^{j-1}\tau)$ so that each lane runs a half-duration evolution in parallel, giving an evolution depth of $U(2^{j-1}\tau)$ rather than $U(2^{j}\tau)$. A detailed resource analysis for this choice is given in Sec.~\ref{sec:resources}.

\paragraph*{Vacuum reference for particle-conserving Hamiltonians.}
For molecular electronic Hamiltonians that conserve particle number, the fermionic vacuum is an especially simple reference. The second-quantized electronic-structure Hamiltonian in normal-ordered form is
\begin{equation}
H_{\mathrm{SQ}} = E_{\mathrm{nuc}} + \sum_{pq} h_{pq}\,a_p^\dagger a_q
+ \frac{1}{2}\sum_{pqrs} h_{pqrs}\,a_p^\dagger a_q^\dagger a_r a_s,
\label{eq:chem_H}
\end{equation}
where $E_{\mathrm{nuc}}$ is a constant shift (typically the nuclear-repulsion energy) and $a_p^\dagger$, $a_q$ are fermionic creation and annihilation operators. The vacuum state $\ket{\mathrm{vac}}$ satisfies $a_q\ket{\mathrm{vac}}=0$ for all $q$, so every excitation term in Eq.~\eqref{eq:chem_H} annihilates $\ket{\mathrm{vac}}$ and $\ket{\mathrm{vac}}$ is an eigenstate of
$H_{\mathrm{SQ}}$.
With typical fermion-to-qubit encodings (e.g., Jordan--Wigner or Bravyi--Kitaev), the vacuum maps to the all-zero computational basis state $\ket{\overline 0}$ \cite{JordanWigner_1928, BravyiKitaev_2002, SeeleyRichardLove_2012}. Therefore,
\begin{equation}
U(\tau)\ket{\overline 0}
=
e^{-i\tau E_{\mathrm{vac}}}\ket{\overline 0}
=
e^{2\pi i \Theta(\tau)}\ket{\overline 0},
\label{eq:vac_phase}
\end{equation}
where $E_{\mathrm{vac}}$ is the classically easily computable vacuum eigenvalue of the Hamiltonian and $\Theta(\tau)= -\frac{\tau E_{\mathrm{vac}}}{2\pi}\ \bmod 1$ is the corresponding vacuum phase. Therefore, when $\ket{\phi}=\ket{\overline 0}$, the estimated phase is $\left(\theta + \Phi_k - \Theta(\tau)\right)\ \bmod 1$, which reduces to the output of canonical QPE with $\theta=\Theta(\tau)$ gadget phase choice.

In practice, after fermion to qubit mapping the qubit Hamiltonian typically contains identity contributions beyond the explicit $E_{\mathrm{nuc}}$. Since in SE-QPE the time evolution, $U(\tau)$ is uncontrolled, the identity terms appear as a net global phase and they become completely unobservable. Therefore, any constant offset to the energy if present relative to the scalar free Hamiltonian must be taken into account via $\theta$ or by classical post-processing.

When a trivial reference eigenstate such as $\ket{\overline 0}$ is used, the reference register also enables a simple error detection method with a reset-based error-suppression. Ideally the reference register is returned to the state $\ket{\overline 0}$ after each gadget application, therefore it can be measured and reset to the exact reference state, as shown in Fig.~\ref{fig:qppe_vacuum_circuit}. While measuring the reference register before mid-circuit reset and/or at the end of a circuit allows error detection by post-selecting on the all-zero outcome, the mid-circuit reset removes other errors accumulated on the reference register and limits their further propagation. Furthermore, the error detection can be executed mid-circuit with early exit in case of failure, saving subsequent runtime.
\paragraph*{Fan-out acceleration of the CSWAP layer.}
Each CSWAP-gadget contains a layer of $N$ controlled-SWAPs, which contributes an $O(N)$ two-qubit depth overhead if implemented with a single control line. To reduce this overhead, one can fan out the control onto an auxiliary register $F$ and execute the controlled-SWAPs in parallel (Fig.~\ref{fig:qppe_vacuum_circuit}) \cite{Huggins_NO_2020}. With a CX-tree construction, the cat (GHZ) state preparation and unpreparation depth scales as $O(\log N)$ in an all-to-all connectivity model \cite{MooreNilsson_2002}. In an ideal noiseless execution, the fan-out register uncomputes back to the all-zero state and can be discarded. In practice, measuring $F$ and post-selecting the all-zero outcome provides an additional error detecting check further to the reference-register check. Like with the reference register this check can occur mid-circuit and be combined with qubit resetting to limit undetected error proliferation.

Throughout the text we refer to variations of the SE-QPE circuits with labels to specify their particular construction. Firstly, QPE refers to a canonical QPE implementation with inverse QFT and controlled-$U(\tau)^{2^j}$ for time evolution. Any 'SE-QPE' circuit refers the split-evolution equivalent of the previously described canonical QPE circuit, for which each time evolution step is performed by an appropriate ${S}_U$ gadget and the reference register is the vacuum state (see Fig.~\ref{fig:qppe_circuit}). The addition of a 'cat' prefix indicates fan-out acceleration of the CSWAP layer via a cat state has been implemented (see Fig.~\ref{fig:qppe_vacuum_circuit}). The acronym 'MR' refers to 'measure and reset' and indicates that the reference register qubits and, if relevant, cat state fan-out qubits (i.e., the $\ket{\overline{0}}_B$ and $\ket{\overline{0}}_F$ registers in Fig.~\ref{fig:qppe_vacuum_circuit}) are measured and reset to the $\ket{\overline 0}$ state mid-circuit, after each ${S}_U$ gadget implementation within a circuit. 'Mixed' SE-QPE circuits are phase-estimation circuits with a mixture of controlled-$U$ blocks and ${S}_U$ gadget time evolution steps.

\begin{figure}[h]
\centering
\begin{quantikz}
\lstick{$\ket{+}_C$} & \qw&  \gategroup[wires=2,steps=5,style={draw,dashed,rounded corners,inner sep=4pt}]{}& \qw& \ctrl{1} & \gate{P(\theta)} & \ctrl{1} & \qw & \qw\\
\lstick{$\ket{\overline 0}_A$} &  \qw&\qw&\qwbundle{N} & \gate{U_{\psi}} & \gate{U} & \gate{U_{\psi}^\dagger} & \qw & \qw
\end{quantikz}
\caption{A compute--uncompute gadget marked with the dashed box, where $U_{\psi}$ is defined as a state preparation unitary on a reference state $\ket{\overline{0}}$ such that it prepares the target state $\ket{\psi} = U_{\psi}\ket{\overline{0}}$.}
\label{fig:compute--uncompute-gadget}
\end{figure}
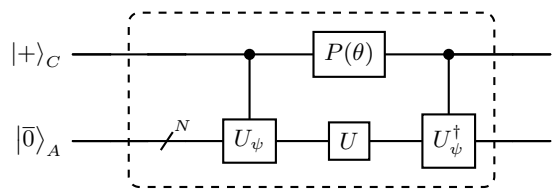

\paragraph*{Compute--uncompute substitution.}
An alternative strategy to avoid controlled time evolution is to control state preparation and unpreparation around an uncontrolled $U(\tau)$ \cite{LinTong_2022}.
Fig.~\ref{fig:compute--uncompute-gadget} shows this substitution of a controlled-$U(\tau)$ with a phase correction on the control. This gadget can provide an effective phase kickback on a specific reference state, but it is not an equivalent substitute for the controlled powers in canonical QPE, except when the initial state is an exact eigenstate. The gadget implements
\begin{equation}
S_U^{\mathrm{CU}}
=
\ket{0}\!\bra{0}\otimes U
+
e^{2\pi i \theta}\ket{1}\!\bra{1}\otimes U_{\psi}^\dagger U U_{\psi},
\label{eq:compute_uncompute_gadget_operator}
\end{equation}
with $P(\theta)=\ket{0}\bra{0}+e^{2\pi i\theta}\ket{1}\bra{1}$. Provided that the reference state $\ket{\overline 0}_A$ is an eigenstate,
\begin{equation}
U(\tau)\ket{\overline 0}_A = e^{2\pi i \Theta(\tau)}\ket{\overline 0}_A,
\end{equation}
and that $U_{\psi}$ prepares an arbitrary state 
\begin{equation}
\ket{\psi}_A = U_{\psi}\ket{\overline 0}_A,
\end{equation}
choosing $\theta=\Theta(\tau)$ yields
\begin{equation}
S_{U}^{\mathrm{CU}} \ket{+}\ket{\overline{0}}
=
e^{2\pi i \Theta(\tau)}
\frac{1}{\sqrt{2}}
\left(
\ket{0}\ket{\overline{0}}
+
\ket{1}\,U_{\mathrm{eff}}(\tau)\ket{\overline{0}}
\right),
\label{eq:CUc_output}
\end{equation}
where $U_{\mathrm{eff}}(\tau)=U_{\psi}^\dagger U(\tau) U_{\psi}$.

If $\ket{\overline{0}}$ is an eigenstate of $U_{\mathrm{eff}}$ (i.e., $\ket{\psi}$ is an eigenstate of $U(\tau)$), then the system register returns to the reference state after the application of $S_{U}^{\mathrm{CU}}$. Therefore, subsequently applied $S_{U^{2^j}}^{\mathrm{CU}}$ is a compatible replacement of the controlled-$U^{2^j}$ in phase estimation.
Conversely, if $\ket{\psi}$ is not an eigenstate of $U(\tau)$, then the system register does not return to the reference state and subsequently applied $S_{U^{2^j}}^{\mathrm{CU}}$ does not yield Eq.~\eqref{eq:CUc_output} as a substitute for controlled-$U^{2^j}$, and the resulting circuit does not implement a canonical QPE equivalent. This inequivalence is demonstrated in Appendix~\ref{app:noiselessCU}. Furthermore, although it has no qubit count overhead, $S_{U^{2^j}}^{\mathrm{CU}}$ requires controlled state preparations, and the serial application of controlled-$U_{\psi}$, $U$, and controlled-$U_{\psi}^\dagger$ could lead to increased runtime overhead compared to SE-QPE.

\section{Gate counts and depths analysis}
\label{sec:resources}

We compare canonical QPE to SE-QPE at the level of blocks that realize the effective controlled powers of $U(\tau)=e^{-i\tau H}$. SE-QPE replaces each controlled-$U(\tau)^{2^j}$ block by the CSWAP-gadget ${S}_{U^{2^j}}$, eliminating controlled time evolution and enabling parallel execution of the evolution operator. The tradeoff is an additive CSWAP-layer overhead per substituted phase bit and the additional qubits for the second register. For resource comparison it is convenient to separate algorithmic scaling from implementation overhead. In particular, we focus only on the time evolution blocks and ignore the cost of the inverse QFT, which is unaffected by any substitution. Furthermore, we consider settings that yield the same outcome distribution as the QPE, using the vacuum reference state, and therefore there is no additional state-preparation overhead. For any circuit block $V$ we track counts of a gate $G$ as $C_{G}[V]$ and corresponding depths $D_{G}[V]$ in the gate set of CX, arbitrary-angle $R_z$, and one-qubit Clifford gates. The corresponding $T$ metrics can be obtained by synthesizing the arbitrary-angle $R_z$ rotations into Clifford+$T$ or using a per-rotation cost $t_\varepsilon$ \cite{ross2016optimal}.

For $U(\tau)$, for both counts and depths of a gate $G$ (e.g., $G=\mathrm{CX}$, $R_z$, or $T$), we define $c_G = C_G\left[U(\tau)\right]$, $d_G =D_G\left[U(\tau)\right]$ and we model the control overhead with  $C_G\left[\mathrm{c}U(\tau)\right] = R_G\,c_G$ and $D_G\left[\mathrm{c}U(\tau)\right] = r_G\,d_G$, where $\mathrm{c}U(\tau)$ denotes the controlled-$U(\tau)$ time evolution and $R_G$ and $r_G$ capture the implementation-dependent overhead of adding controls. The CSWAP-gadget adds an overhead per substituted phase bit and we denote this per-bit overhead as $c_G^{\mathrm{swap}}$, $d_G^{\mathrm{swap}}$.

In canonical QPE, the $j$-th phase-kickback block is realized by repeating the controlled base step $2^j$ times:
\begin{equation}
  C_G^{(j),\mathrm{QPE}} = 2^j\,R_G\,c_G,
  \qquad
  D_G^{(j),\mathrm{QPE}} = 2^j\,r_G\,d_G.
  \label{eq:block_qpe}
\end{equation}
In SE-QPE, the same block is implemented by one gadget ${S}_{U^{2^j}}$ and uncontrolled evolution on two lanes.
For $j>0$ we use the balanced choice $U_A=U_B=U(\tau)^{2^{j-1}}$, so that each lane runs half the evolution time in parallel.
This preserves the total number of base-step evolutions across both lanes but halves the evolution depth, that is
\begin{equation}
  \begin{split}
  C_G^{(j),\mathrm{SE-QPE}} = 2^j\,c_G + c_G^{\mathrm{swap}},
  \quad
  \\
  D_G^{(j),\mathrm{SE-QPE}} = 2^{j-1}\,d_G + d_G^{\mathrm{swap}}.
  \label{eq:block_qppe}
  \end{split}
\end{equation}
For $j=0$, balanced halving requires implementation-specific access to $U(\tau)$ or a smaller time step. If we choose the completely unbalanced factorization then $C_G^{(0),\mathrm{SE-QPE}} = c_G + c_G^{\mathrm{swap}}$ and $D_G^{(0),\mathrm{SE-QPE}} = d_G + d_G^{\mathrm{swap}}$. By comparison, for bit $j$ the substitution is beneficial for counts when
\begin{equation}
  2^j\,(R_G-1)\,c_G > c_G^{\mathrm{swap}},
  \label{eq:block_breakeven_counts}
\end{equation}
and beneficial for depths (for $j>0$) when
\begin{equation}
  2^{j-1}\,(2r_G-1)\,d_G > d_G^{\mathrm{swap}}.
  \label{eq:block_breakeven_depth}
\end{equation}
For simple systems with small $j$, the CSWAP layer overhead can dominate and a direct controlled implementation may still be preferable, while for large $j$ the multiplicative overhead of controlled time evolution dominates and the substitution becomes favorable. This motivates mixed strategies in which only higher-$j$ blocks are substituted.

By summing over all $M$ bits, the canonical QPE costs are (without QFT and state preparation)
\begin{equation}
  C_G^{\mathrm{QPE}} \simeq K\,R_G\,c_G,
  \qquad
  D_G^{\mathrm{QPE}} \simeq K\,r_G\,d_G
  \label{eq:total_qpe_rc}
\end{equation}
and the SE-QPE costs are
\begin{equation}
  \begin{split}
  C_G^{\mathrm{SE-QPE}} \simeq K\,c_G + M\,c_G^{\mathrm{swap}},
  \\
  D_G^{\mathrm{SE-QPE}} \simeq \frac{K+1}{2}\,d_G + M\,d_G^{\mathrm{swap}},
  \label{eq:total_qppe_rc}
  \end{split}
\end{equation}
resulting in net gain ratios in the time evolution portion of the phase-estimation circuits of
\begin{equation}
g_G^{\mathrm{count}}=\frac{C_G^{\mathrm{SE-QPE}}}{C_G^{\mathrm{QPE}}}
  \simeq
  \frac{1}{R_G}
  \left(
    1+\frac{M}{K}\frac{c_G^{\mathrm{swap}}}{c_G}
  \right),
  \label{eq:gain_counts_rc}
\end{equation}
and
\begin{equation}
g_G^{\mathrm{depth}}=\frac{D_G^{\mathrm{SE-QPE}}}{D_G^{\mathrm{QPE}}}
  \simeq
  \frac{1}{2r_G}
  \left(
    1+\frac{1}{K}+\frac{2M}{K}\frac{d_G^{\mathrm{swap}}}{d_G}
  \right)
  \label{eq:gain_depth_rc}
\end{equation}
The leading improvements are set by the control overheads $R_G$, $r_G$ and the additional factor of $1/2$ in depth from two-lane parallelization. The CSWAP overhead enters only through the factor $M/K$, which decays rapidly. 

Similar substitution gains apply to iterative QPE. Each iteration requires a controlled application of $U(\tau)^{2^j}$ and this controlled block can be replaced by the CSWAP-gadget ${S}_{U^{2^j}}$ (either in a SE-QPE or cat-SE-QPE implementation of the gadget) with the same balanced factorization as above. Consequently, the per-iteration costs follow Eqs.~\eqref{eq:block_qpe} and \eqref{eq:block_qppe}.

Because controlling a black-box unitary is not an explicit circuit-level operation, the control overhead must be evaluated for a specific implementation\cite{Araujo_2014}. We therefore consider Trotterized implementations of time evolution for the second-quantized Hamiltonian in double-factorized form  \cite{Motta_2021, PhysRevLett.120.110501},
\begin{equation}
  \hat H = \hat W_0\!\left(\sum_{i=0}^{N-1}\alpha_i Z_i\right)\hat W_0^\dagger
  + \sum_{l=1}^{L}\hat W_l\!\left(\sum_{i>j}\beta_{ij}^{(l)} Z_i Z_j\right)\hat W_l^\dagger,
\end{equation}
where $L$ is the number of double-factorized terms and $\hat W_l
  = e^{\sum_{i>j}[\log \mathbf{W}_l]_{ij}(a^{\dagger}_ia_j -  a^{\dagger}_ja_i)}$
are basis rotations that transform the one- and two-body blocks into $Z$- and $ZZ$-diagonal form \cite{Motta_2021, PhysRevLett.120.110501}. The substitutions can be used with alternative implementations as well, not only with Trotterization. In the analytical cost estimates we assume $\alpha_i \neq 0$ and $\beta_{ij}^{(l)} \neq 0$, however in practice, some coefficients may be close to zero. Using single-step first- or second-order Trotterization, the time evolution operator can
be taken as
\begin{equation}
  U^{(1)}(\tau) =
  \hat W_0\Bigg[
    U_0(\tau)
    \prod_{l=1}^L  \hat W_{l-1,l}\,U_l(\tau)
  \Bigg]\hat W_L^\dagger,
  \label{eq:trotter_U}
\end{equation}
or, using the second-order symmetric formula,
\begin{equation}
  \begin{split}
    U^{(2)}(\tau) =
    \hat W_0\Bigg[
      U_0(\tau/2)
      \left(\prod_{l=1}^{L-1} \hat W_{l-1,l}\,U_l(\tau/2)\right) \\
      \times \hat W_{L-1,L}\,U_L(\tau)
      \left(\prod_{l=L}^{1} \hat W_{l,l-1}\,U_{l-1}(\tau/2)\right)
    \Bigg]
\hat W_0^\dagger,
  \end{split}
  \label{eq:trotter_U2}
\end{equation}
where
\begin{equation}
  U_0(\tau) = e^{-i\tau\sum_i \alpha_i Z_i},
  \qquad
  U_l(\tau) = e^{-i\tau\sum_{i>j}\beta^{(l)}_{ij} Z_i Z_j},
\end{equation}
and $\hat W_{l-1,l} = \hat W_{l-1}^{\dagger}\hat W_{l}$, with $\hat W_{l,l-1}
= \hat W_l^\dagger \hat W_{l-1} = \hat W_{l-1,l}^\dagger$ and the product $\prod_{l=L}^{1}$ is ordered with decreasing $l$.
Since $\hat W_{l-1,l}$ (and $\hat W_{l,l-1}$) has the same circuit structure as any
other $\hat W_l$, we cost all such blocks using a single representative. In what follows we write $U(\tau)$ to denote either $U^{(1)}(\tau)$ or $U^{(2)}(\tau)$. The costs for the building blocks $W_l$, $U_0(\tau)$, $U_l(\tau)$, their controlled versions, and the CSWAP layers used are summarized in Table~\ref{tab:primitive_costs}. In particular, under all-to-all connectivity and even $N$, the dense $ZZ$ kernels $U_l(\tau)$ can be scheduled in $O(N)$ two-qubit depth via perfect matchings \cite{Behzad_1967,Guerreschi_2018}, whereas their controlled counterparts are ancilla-bottlenecked and scale as $O(N^2)$ in two-qubit depth (Table~\ref{tab:primitive_costs}).

\begin{table*}[t]
\centering
\renewcommand{\arraystretch}{1.25}
\setlength{\tabcolsep}{4pt}
\scriptsize
\begin{tabular*}{\textwidth}{@{\extracolsep{\fill}}c|ccccccc}
\hline\hline
 Gate  & $W_l$ & $U_0(\tau)$ & $U_l(\tau)$ & $\mathrm{c}U_0(\tau)$ & $\mathrm{c}U_l(\tau)$ & CSWAP ladder pair & cat-CSWAP ladder pair \\
 Metrics &  &  & ($l>0$) &  & ($l>0$) & (serial) & (fan-out + parallel) \\
\hline
$C_{\mathrm{CX}}$
& $N(N-1)$
& $0$
& $N(N-1)$
& $2N$
& $2N(N-1)$
& $14N$
& $14N + 2(N-1)$
\\
$D_{\mathrm{CX}}$
& $4N-6$
& $0$
& $2(N-1)$
& $2N$
& $(N+2)(N-1)$
& $14N$
& $14 + 2\lceil \log_2 N\rceil$
\\
$C_{R_z}$
& $N(N-1)$
& $N$
& $\frac{N(N-1)}{2}$
& $2N$
& $N(N-1)$
& -
& -
\\
$D_{R_z}$
& $2N-3$
& $1$
& $N-1$
& $2$
& $2(N-1)$
& -
& -
\\
$C_T$
& $N(N-1)t_\varepsilon$
& $Nt_\varepsilon$
& $\frac{N(N-1)}{2}t_\varepsilon$
& $2Nt_\varepsilon$
& $N(N-1)t_\varepsilon$
& $14N$
& $14N$
\\
$D_T$
& $(2N-3)t_\varepsilon$
& $t_\varepsilon$
& $(N-1)t_\varepsilon$
& $2t_\varepsilon$
& $2(N-1)t_\varepsilon$
& $8N$
& $8$
\\
\hline\hline
\end{tabular*}
\caption{
Resource costs for the building blocks used in Eq.~\eqref{eq:trotter_U}.
For $U_l(\tau)$, assuming all-to-all connectivity on an even number of $N$ qubits, the $N(N-1)/2$ distinct $Z_i Z_j$ pairs can be decomposed into $N-1$ disjoint perfect matchings, each containing $N/2$ pairs %
\cite{Behzad_1967,Guerreschi_2018,kivlichan2020improved}.
For CSWAP ladder-pair columns we use an all-to-all decomposition with $7$ CXs per CSWAP gate %
\cite{mq2024shallow}, $T$-count $7$%
~\cite{gosset2013algorithm}
and $T$-depth $4$%
~\cite{park2025reducing}.
The cat-CSWAP layer pair includes fan-out+unfan-out of the control into a GHZ/cat state using a CX tree with $2(N-1)$ CXs and depth $2\lceil\log_2 N\rceil$ %
\cite{prabhu2021fault}.
Here $t_\varepsilon$ is the fault-tolerant $T$-count or $T$-depth per synthesized arbitrary-angle $R_z$ at precision $\varepsilon$ %
\cite{ross2016optimal,kliuchnikov2013asymptotically}.
}
\label{tab:primitive_costs}
\end{table*}

In the case of controlled-$U(\tau)$, the basis rotations do not need to be controlled. When the control is $\ket{0}$ the diagonal kernels act as the identity, and the remaining uncontrolled basis rotations cancel to identity \cite{Cowtan2020}. Therefore only the diagonal terms need to be controlled, as $\exp\!\left(-i\theta |1\rangle\!\langle 1|_a Z_i\right)$ and $\exp\!\left(-i\theta |1\rangle\!\langle 1|_a Z_iZ_j\right)$. For the analytical gate counts and depths we used the identities $e^{-i\theta Z_i} = R_z^{(i)}(2\theta)$, $e^{-i\theta Z_i Z_j}
  =
  \mathrm{CX}_{i\to j}\, R_z^{(j)}(2\theta)\, \mathrm{CX}_{i\to j}$, $e^{-i\theta \ketbra{1}{1}_a Z_i}
  =
  R_z^{(i)}(\theta)\,
  \mathrm{CX}_{a\to i}\,
  R_z^{(i)}(-\theta)\,
  \mathrm{CX}_{a\to i}$ and $e^{-i\theta \ketbra{1}{1}_a Z_i Z_j}
  =
  \mathrm{CX}_{i\to j}\,
  R_z^{(j)}(\theta)\,
  \mathrm{CX}_{a\to j}\,
  R_z^{(j)}(-\theta)\,
  \mathrm{CX}_{a\to j}\,
  \mathrm{CX}_{i\to j}$ and for the nearest-neighbour Givens rotation\\ $\hat G_{p,p+1}(\theta)
  =
  H_p\,
  \mathrm{CX}_{p\to p+1}\,
  R_y^{(p)}(-\theta)\,R_y^{(p+1)}(-\theta)\,
  \mathrm{CX}_{p\to p+1}\,
  H_p,$
with $R_y^{(q)}(\theta)=e^{-i\frac{\theta}{2}Y_q}$.

The substitution gadgets require two layers of an $N$ CSWAP cascade between the two system registers.
We cost one CSWAP as 7 CXs with two-qubit depth 7 \cite{mq2024shallow}. A series of $N$ CSWAPs thus contributes $7N$ CXs and depth $7N$, so a pair of them used in the gadget contributes $14N$ CXs and depth $14N$ per substituted phase bit. In a simple cat state variant, the shared control is fanned out into $N$ controls so the $N$ CSWAPs can run in parallel at the expense of $N-1$ extra qubits and the cat state preparation and unpreparation layers.

The total per-Trotter-step counts and depth, together with the control overhead ratios, are in Table~\ref{tab:step_costs_ratios_overheads_compact} and Table~\ref{tab:step_depths_ratios_overheads_compact}, respectively. By combining the costs of the building blocks we find the resource reduction gains in the limits as
\begin{equation}
g_{\mathrm{CX}}^{\mathrm{count}}\simeq\frac{2}{3},\qquad g_{\mathrm{CX}}^{\mathrm{depth}}\simeq\frac{3}{N}
\label{eq:gain_factors_cx}
\end{equation}
and
\begin{equation}
g_{\mathrm{R_z}}^{\mathrm{count}}\simeq\frac{3}{4},\qquad g_{\mathrm{R_z}}^{\mathrm{depth}}\simeq\frac{3}{8}.
\label{eq:gain_factors_rz}
\end{equation}
In near-term simulations, the dominant limitation is often two-qubit error accumulation and depth, so $D_{\mathrm{CX}}$ or $g_{\mathrm{CX}}^{\mathrm{depth}}$ are particularly relevant figures of merit. The removal of controlled diagonal kernels and the two-lane parallelization directly reduce the required two-qubit depth per phase bit encoding. In the fault-tolerant setting the bottleneck typically shifts to $T$ resources, and assuming an approximately constant synthesis cost $t_\varepsilon$ for arbitrary-angle $R_z$ rotations, the $R_z$ ratios translate directly to $T$-count and $T$-depth gains. The depth scalings used in Table~\ref{tab:primitive_costs} assume all-to-all connectivity (e.g., trapped-ion architectures) so that dense $ZZ$ layers can be scheduled via perfect matchings and cat state fan-out can be implemented in $O(\log N)$ depth. On architectures with restricted connectivity, additional routing overhead will modify both the baseline evolution and the CSWAP-layer costs, and the substitution gains should be reevaluated using architecture-specific constraints.

\begin{table}[t]
\centering
\scriptsize
\renewcommand{\arraystretch}{1.25}
\setlength{\tabcolsep}{6pt}
\begin{tabular*}{\columnwidth}{@{\extracolsep{\fill}}l|cc}
\hline\hline
 $G$ & $\mathrm{CX}$ & $R_z(\theta)$ \\
\hline
$C_G[U^{(1)}(\tau)]$ &
$2LN^2+2N^2-2LN-2N$ &
\makecell[c]{$\frac{3}{2}LN^2+2N^2-$\\$\frac{3}{2}LN-N$}
\\\hline
$C_G[\mathrm{c}U^{(1)}(\tau)]$ &
$3LN^2+2N^2-3LN$ &
$2LN^2+2N^2-2LN$
\\\hline
$C_G[U^{(2)}(\tau)]$ &
$4LN^2+N^2-4LN-N$ &
\makecell[c]{$3LN^2+\frac{3}{2}N^2+$\\$-3LN+\frac{1}{2}N$}
%\substack{$LN^2-\frac{1}{2}N^2+$&\\$+3LN+\frac{13}{2}N$} &
\\\hline
$C_G[\mathrm{c}U^{(2)}(\tau)]$ &
$6LN^2-6LN+4N$ &
\makecell[c]{$4LN^2+N^2+$\\$-4LN+3N$}
\\\hline
$R_G$ &
$\simeq \frac{3}{2}$ &
$\simeq \frac{4}{3}$
\\
\hline\hline
\end{tabular*}
\caption{Per-Trotter-step costs and the control overhead ratios
$R_G = \frac{C_G[\mathrm{c}U(\tau)]}{C_G[U(\tau)]}$,
corresponding to gate counts evaluated at leading order with $L>N$. For both Trotter orders, the leading-order scaling is identical.
}
\label{tab:step_costs_ratios_overheads_compact}
\end{table}

\begin{table}[t]
\centering
\scriptsize
\renewcommand{\arraystretch}{1.25}
\setlength{\tabcolsep}{6pt}
\begin{tabular*}{\columnwidth}{@{\extracolsep{\fill}}l|cc}
\hline\hline
 $G$ & $\mathrm{CX}$ & $R_z(\theta)$ \\
\hline
$D_G[U^{(1)}(\tau)]$ &
$6LN+8N-8L-12$ &
$3LN+4N-4L-5$
\\\hline
$D_G[\mathrm{c}U^{(1)}(\tau)]$ &
$LN^2+5LN+10N-8L-12$ &
$4LN+4N-5L-4$
\\\hline
$D_G[U^{(2)}(\tau)]$ &
$12LN+6N-16L-10$ &
$6LN+3N-8L-3$
\\\hline
$D_G[\mathrm{c}U^{(2)}(\tau)]$ &
\makecell[c]{$2LN^2-N^2+10LN+$\\$+11N-16L-10$} &
$8LN-10L+2N$
\\\hline
$r_G$ &
$\simeq \frac{N}{6}$ &
$\simeq \frac{4}{3}$
\\
\hline\hline
\end{tabular*}
\caption{Per-Trotter-step costs and the control overhead ratios
$r_G = \frac{D_G[\mathrm{c}U(\tau)]}{D_G[U(\tau)]}$,
corresponding to circuit depth evaluated at leading order with $L>N$. For both Trotter orders, the leading-order scaling is identical.
}
\label{tab:step_depths_ratios_overheads_compact}
\end{table}

\section{FeMoco double-factorized Trotterization resource estimates}
\label{sec:femoco_resources}

\begin{figure}[t]
\centering
\includegraphics[width=0.48\textwidth]{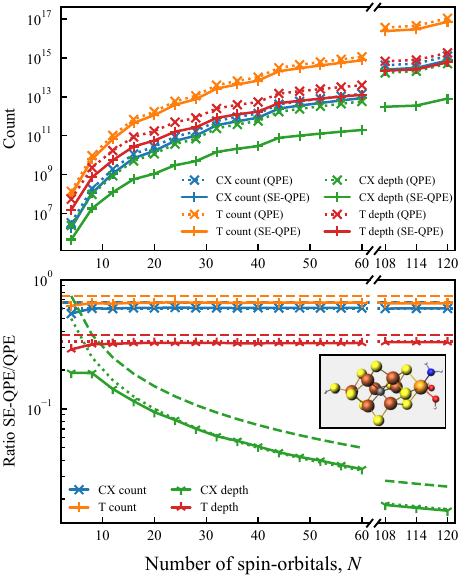}
\caption{FeMoco resource-estimation summary from compiled second-order Trotterization. Top panel shows total CX count, T count, CX depth, and T depth for QPE (dashed) and SE-QPE (solid) versus spin-orbitals $N$. Bottom panel shows the corresponding ratios of SE-QPE to QPE, with a small inset showing the FeMoco geometry. Dashed curves indicate the asymptotic gain factors (Eqs.~\eqref{eq:gain_factors_cx} and \eqref{eq:gain_factors_rz}), dotted curves indicate asymptotic gain factors assuming spin-block-diagonal structure of $W$. Here $\tau=\pi/\Lambda_L$, $n_{\mathrm{Trot}}$ is set by the second-order Trotter heuristic, and cat-CSWAP overhead is included. 
}
\label{fig:femoco_resource_from_csv}
\end{figure}

Using the derived resource estimation formulas, we applied them to a realistic system, evaluating FeMoco across a range of active spaces (AS). FeMoco is the
cofactor of the nitrogenase enzyme that performs biological nitrogen fixation
under mild conditions (room temperature and ambient pressure) %
\cite{hoffman2014mechanism}.
It is a standard test system for quantum-advantage studies in electronic
structure, since it contains multiple Fe atoms and one Mo atom and requires
large active spaces to capture its correlated electronic structure %
\cite{Reiher_2017,li2019electronic,montgomery2018strong,lee2021even}.
Here we report resource metrics across a range of FeMoco active-space
Hamiltonians using a double-factorized (DF) representation and Trotterized time
evolution %
\cite{Reiher_2017,li2019electronic}.
To obtain the DF Hamiltonian used in Sec.~\ref{sec:resources}, we carried out a
restricted Hartree--Fock calculation for the neutral FeMoco model from
Ref ~\cite{montgomery2018strong} using PySCF %.
\cite{sun2018pyscf,sun2020recent}.
We used the crenbl basis and crenbl effective core potential (ECP) for Fe and
Mo %
\cite{fernandez1985ab,ermler1991ab},
and the 6-31G** basis set for the remaining atoms. The model contains a
MoFe$_7$S$_9$C scaffold (inset of Fig.~\ref{fig:femoco_resource_from_csv}). Using the Hartree--Fock
orbital basis, we generated the second-quantized Hamiltonian and then applied
double factorization of the two-electron integrals to obtain the coefficients
$\alpha_i$ and $\beta_{ij}^{(l)}$ using our software suite InQuanto %
\cite{inquanto}.
Hamiltonians were generated for AS ranging from (2e,2o) up to (30e,30o) and (54e,54o), (57e,57o) and (60e,60o).

To provide a simple resource assessment, four sources of error were
considered with equal weights: truncation error, Trotterization error,
$T$-synthesis error, and phase-estimation energy-grid error.
Resources are reported for phase-estimation at target chemical accuracy
$\varepsilon_{\mathrm{chem}}=1.6\times 10^{-3}$ Ha, with the equal split of
\begin{equation}
\varepsilon_{\mathrm{trunc}}=\varepsilon_{\mathrm{TS}}=\varepsilon_{\mathrm{synth}}=\varepsilon_{\mathrm{grid}}=\varepsilon_{\mathrm{chem}}/4 .
\end{equation} Here $\varepsilon_{\mathrm{TS}}$ denotes the energy-error budget assigned to the second-order Trotter-Suzuki approximation. Truncation error is used to determine the smallest $L$ for which the discarded Hamiltonian coefficient-norm bound is below $\varepsilon_{\mathrm{trunc}}$. We define the
retained coefficient-norm bound as
\begin{equation}
\Lambda_L
=
\sum_{i=0}^{N-1}\left|\alpha_i\right|
+\sum_{l=1}^{L}\sum_{i>j}\left|\beta_{ij}^{(l)}\right|.
\label{eq:femoco_LambdaH}
\end{equation}
In this section we set the evolution time for phase-estimation to
\begin{equation}
\tau = \pi/\Lambda_L ,
\end{equation}
and the number of phase bits is chosen from the grid condition
\begin{equation}
M=\left\lceil \log_2\!\left(\frac{\pi}{\varepsilon_{\mathrm{grid}}\tau}\right) \right\rceil + 1 ,
\end{equation}
with the total number of phase-kickback blocks given by $K = 2^M - 1$.

For accurate time evolution we extend the per-step costs in
Sec.~\ref{sec:resources} to a multi-step second-order Trotterized simulation of
the form
\begin{equation}
U(\tau) = U^{(2)}(\tau/n_{\rm Trot})^{n_{\rm Trot}} \approx e^{-i\tau H},
\end{equation}
where $U^{(\mathrm{ord.})}(\tau/n_{\rm Trot})$ denotes a single DF-Trotter step
as in Eqs.~\eqref{eq:trotter_U} or~\eqref{eq:trotter_U2}. For the second-order
Trotterization the Trotter number is estimated with a heuristic %
\begin{equation}
n_{\mathrm{Trot}} \propto \frac{\,\tau\,N^{\xi}}{\sqrt{\varepsilon_{\mathrm{TS}}}}  ,
\end{equation}
with $\xi=1.7$ \cite{Poulin2015TrotterStep}. Further details are given in Appendix~\ref{app:femoco_l_scaling}.

%where $\xi=1.7$ and the scaling constant $c_{\mathrm{TS}}$ is calibrated from
%the reference Trotter error
%$\delta_{\mathrm{ref}}=\left\|e^{-i\tau_{\mathrm{ref}} H} - U^{(2)}(\tau_{\mathrm{ref}})\right\|$
%computed for AS (6e,6o), with $N_{\mathrm{ref}}=12$ at
%$\tau_{\mathrm{ref}} = 1~\mathrm{Ha}^{-1}$, for which we obtain the numerical value 
%$\delta_{\mathrm{ref}}=0.01630246995524085$. T
%$\delta_{\mathrm{ref}}=0.01630$ and 
%\begin{equation}
%c_{\mathrm{TS}}=\frac{\sqrt{\delta_{\mathrm{ref}}/\tau_{\mathrm{ref}}^{3}}}{N_{\mathrm{ref}}^{\xi}} \approx 0.00187
%\end{equation} in the corresponding units \cite{Poulin2015TrotterStep}.

We compare canonical QPE to SE-QPE at the level of the time evolution portion
(QFT costs are omitted since they are identical). In
Eqs.~\eqref{eq:total_qpe_rc} and \eqref{eq:total_qppe_rc} the per-step
quantities change to
$c_G = n_{\mathrm{Trot}}C_G[U^{(2)}(\tau/n_{\rm Trot})]$,
$d_G = n_{\mathrm{Trot}}D_G[U^{(2)}(\tau/n_{\rm Trot})]$ and
$R_Gc_G = n_{\mathrm{Trot}}C_G[\mathrm{c}U^{(2)}(\tau/n_{\rm Trot})]$,
$r_Gd_G = n_{\mathrm{Trot}}D_G[\mathrm{c}U^{(2)}(\tau/n_{\rm Trot})]$.
For AS up to (30e,30o) we compiled the circuits of
$\mathrm{c}U^{(2)}(\tau_{\mathrm{ref}})$ and $U^{(2)}(\tau_{\mathrm{ref}})$ with
\texttt{pytket} \cite{pytket_paper}, while for larger AS we only compiled the representative parts of the circuits discussed in Sec. \ref{sec:resources}. The pass sequence and counting procedure are given in Appendix~\ref{app:femoco_l_scaling}. The resulting gate counts and depths are showed in
Fig.~\ref{fig:femoco_resource_from_csv}.

To obtain $T$ counts, we estimate the cost of synthesizing arbitrary-angle
$R_z$ rotations using
\begin{equation}
\varepsilon_{\mathrm{rot}}=\frac{\tau\,\varepsilon_{\mathrm{synth}}}{\widetilde C_{R_z}},
\end{equation}
where $\widetilde C_{R_z}=R_{R_z}c_{R_z}$ for QPE and $\widetilde C_{R_z}=c_{R_z}$ for SE-QPE, and
\begin{equation}
t_{\varepsilon}=3\log_2\!\left(\frac{1}{\varepsilon_{\mathrm{rot}}}\right)+10 
\end{equation} per-rotation T-count with the per-rotation error \cite{Selinger2015, ross2016optimal}, from this we get the total T-count as
\begin{equation}
T_{\mathrm{tot}}^{(R_z)} = C_{R_z}^{\mathrm{tot}}\,t_{\varepsilon},
\end{equation}
where $C_{R_z}^{\mathrm{tot}}$ is either $C_{R_z}^{\mathrm{QPE}}$ or
$C_{R_z}^{\mathrm{SE-QPE}}$, and in the case of SE-QPE we add the explicit
Clifford+$T$ cost of the CSWAP layers when reporting SE-QPE totals
(Table~\ref{tab:primitive_costs}).

Fig.~\ref{fig:femoco_resource_from_csv} shows that the CX and T gate counts and depths for SE-QPE are consistently lower than those of QPE. The resource reduction from SE-QPE is most significant with respect to gate depth and greater for CX gates than T gates, as expected from the leading analytic relations Eqs.~\eqref{eq:gain_factors_cx} and \eqref{eq:gain_factors_rz}. For each metric the bottom panel on Fig.~\ref{fig:femoco_resource_from_csv} illustrates that beyond $N \approx10$ the gain ratios approach their leading behavior, where the CSWAP layer overhead becomes insignificant. Beyond $N \approx10$ gain ratios remain constant for all included metrics besides CX depth which favorably decreases, due to the assumed all-to-all connectivity and perfectly parallelized scheduling.

\section{Simulations and Hardware Demonstration}
\label{sec:experiments}
To demonstrate the substitution mechanism, parallelism, resource counts and
error detection features of SE-QPE on a chemistry-motivated example we apply it to a
small chemistry-inspired model, the Pariser--Parr--Pople (PPP) $\pi$-electron, 4-spin-orbital
Hamiltonian for ethylene in the $(N_\uparrow,N_\downarrow)=(1,1)$ particle sector. For our test systems we
find that SE-QPE reproduces the canonical QPE phase-register statistics under a
vacuum reference, using an optimized Trotterization. We present a proof of principle hardware
implementation up to 8 bits of precision on the Quantinuum trapped-ion device H2-2 \cite{H2-2}, also demonstrating reference-register
and cat-register mid-circuit reset and error detection.

The PPP Hamiltonian maps to a 4-qubit model after Jordan--Wigner encoding as
\begin{equation}
H = H_1 + H_2,
\end{equation}
with
\begin{equation}
H_1 = \alpha\,(X_{1}X_{2}+Y_{1}Y_{2}+X_{3}X_{4}+Y_{3}Y_{4}),
\end{equation}
and
\begin{equation}
H_2 = \beta_{1}\,(Z_{1}Z_{4}+Z_{2}Z_{3})
+ \beta_{2}\,(Z_{1}Z_{3}+Z_{2}Z_{4}),
\end{equation}
where the numerical values representing the energetics of the ethylene \cite{zhang2011excitation} are
$\alpha = -0.055557~\mathrm{Ha}$,
$\beta_{1} = 0.067525~\mathrm{Ha}$,
$\beta_{2} = 0.104616~\mathrm{Ha}$,
and the vacuum reference state is $\ket{\overline 0}=\ket{0000}$. We prepare the
input state using the 4-qubit singlet ansatz $\ket{\psi(\vartheta)}$ shown in
Fig.~\ref{fig:ansatz_state}. The mean-field state is obtained by
$\vartheta_{\mathrm{MF}}=\pi/4$ and has overlap
\begin{equation}
|\braket{S_0}{S_0^{\mathrm{MF}}}|^2 = 0.97427,
\end{equation}
with the exact singlet ground-state $\ket{S_0}$
(Appendix~\ref{app:ethylene_model}). In canonical QPE, this overlap sets the dominant peak
weight in the phase distribution. In SE-QPE with a vacuum reference the same
peak weights are reproduced in the phase register (Sec.~\ref{sec:pcpe}).

For the experiments we implement the base-step unitary, $U(\tau)$, using a single-step symmetric
product
\begin{equation}
e^{-i\tau H} \approx
e^{-i s_1 H_1}e^{-i s_2 H_2}e^{-i s_1 H_1}.
\end{equation}
As detailed in Appendix~\ref{app:trotter_ethylene}, the vacuum $\ket{\overline 0}$ satisfies
$H_1\ket{\overline 0}=0$, so the reference phase depends only on $H_2$. We
therefore set
\begin{equation}
s_2=\tau,
\end{equation}
so that the reference-state phase is reproduced exactly at evolution time
$\tau$, independently of $s_1$.

To reduce the leading Trotter bias on the ground-state phase we take
\begin{equation}
s_1=\frac{\tau}{2}+\lambda\tau^3,
\end{equation}
where $\lambda$ is chosen using a simple perturbative condition for an
available approximate ground-state $\ket{\psi}$, as derived in
Appendix~\ref{app:trotter_ethylene}. For the mean-field state $\ket{S_0^{\mathrm{MF}}}$ we
obtain $\lambda \approx 0.00091716~\mathrm{Ha}^2$. 

We first validate the expected QPE distribution using exact statevector
simulation (SV), for $M=5$ and $M=7$ phase qubits with $\tau=10~\mathrm{Ha}^{-1}$ and $\tau=8~\mathrm{Ha}^{-1}$, respectively, the QPE distributions obtained from the mean-field input states are shown in
Fig.~\ref{fig:statevector}. These time steps are chosen to balance phase-estimation readout with Trotter error (see Appendix \ref{app:timestep}). In the top panels of Fig. \ref{fig:statevector} the black curves show the effect of Trotterization and the reduction of the Trotter bias on the simple SV based QPE probabilities. The dashed black curves show the probabilities for the Trotter error free exact time evolution operator, while the dotted black curves show the probabilities for the single-step second-order Trotterized time evolution operator. The solid black curves show the Trotter bias reduced results with the calculated $\lambda$ value, showing that distribution peaks aligned well with those of the Trotter free results.

In both, $M=5$ and $M=7$  cases the dominant peak corresponds to the singlet ground-state and the secondary peak to an excited singlet state.
For $M=5$ and $M=7$ the estimated energies from the most likely bitstrings are $\widetilde{E}_{GS}^{(\mathrm{SV})} = -0.235619~\mathrm{Ha}$ and
$\widetilde{E}_{GS}^{(\mathrm{SV})} = -0.239301~\mathrm{Ha}$, respectively,
which are within $1.35~\mathrm{mHa}$ and $5.02~\mathrm{mHa}$ of the exact ground-state energy
$E_{GS}=-0.234282~\mathrm{Ha}$.

As the resolution of the computed energy estimates increases with the number of phase qubits a difference between the exact evolution and the trotterized statevector peak outcomes becomes apparent. Hence, the energy estimate calculated with the $M=7$ statevector QPE simulation appears further from the exact ground-state value than with that of $M=5$.

To numerically demonstrate that SE-QPE reproduces the phase-register statistics of QPE, the top panel histograms of Fig.~\ref{fig:statevector} show the noiseless simulation result of the SE-QPE circuits on the Quantinuum H2-1LE simulator \cite{H2-1LE}. In the bottom panels we plot the output of hardware experiments for $M=5$ and $M=7$ MR-cat-SE-QPE circuits, in addition to a $M=7$ cat-SE-QPE circuit all run on the Quantinuum H2-2 quantum computer \cite{H2-2} using the same initial target state on register
$A$ and the vacuum reference $\ket{\overline 0}$ on register $B$. In these (and all subsequent) SE-QPE-based circuits we
set the gadget phase correction, $\theta=\Theta(\tau)$, to cancel the reference phase,
where $\Theta(\tau)$ is computed from the vacuum energy $E_{\mathrm{vac}}=0.344282~\mathrm{Ha}$. Inspecting the
plots, we find agreement in the dominant peak locations.

\begin{figure*}[t]
\centering
\includegraphics[width=0.98\textwidth]{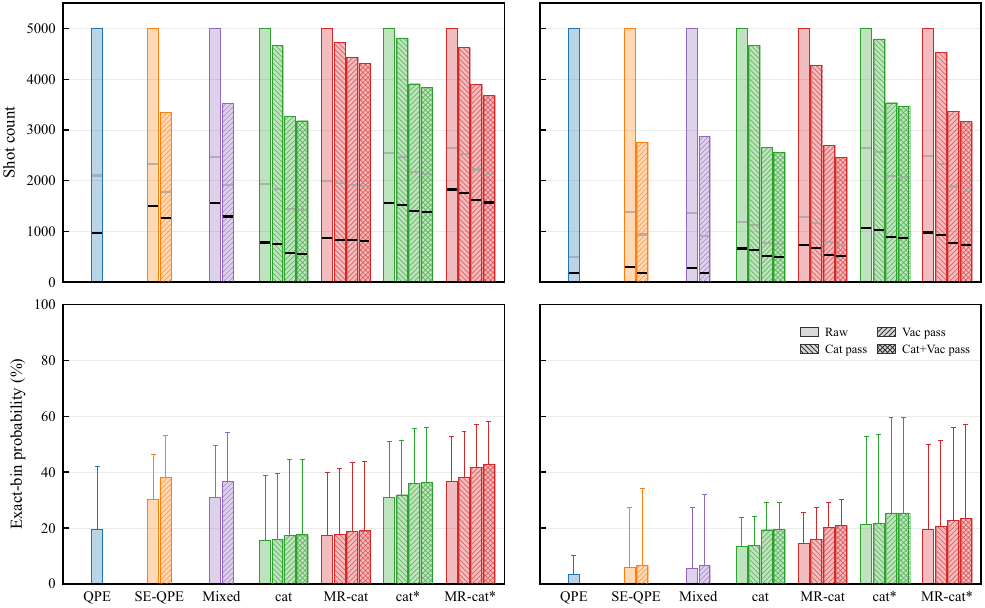}
\caption{(Top) The proportion of shots which returned the modal outcome for variants of SE-QPE and QPE in shot-based emulator and hardware experiments (described in Table \ref{tab:ethylene_result}). With $\tau=10$ Ha$^{-1}$ for $M=5$ (left) and $\tau=8$ Ha$^{-1}$ for $M=6$ (right). For SE-QPE variants results filtered by all-zero fan-out and/or reference register measurements are shown beside the raw data and differentiated by hatched patterns. Results that pass fan-out qubit checks are hatched top left to bottom right and results that pass reference register checks are hatched from top right to bottom left. (Bottom) Plots showing the probability any result falls within the exact 5-bit (left) or 6-bit (right) phase bins for a selection of the emulator and hardware experiments in Table \ref{tab:ethylene_result}. Each box's whiskers encompass the probabilities of one bin either side of the exact phase. The $^*$ denotes real hardware data \cite{H2-2}.}
\label{fig:abs_rel}
\end{figure*}

\emph{Noisy simulations and error detection filtering.}
We perform noisy simulations of QPE and SE-QPE variants, for $M=5$ phase qubits with $\tau=10~\mathrm{Ha}^{-1}$ and $M=6$ phase qubits with $\tau=8~\mathrm{Ha}^{-1}$, using the Quantinuum H2-Emulator with noise models mirroring a real System Model H2 device \cite{H2-Emulator}. The results are summarized in Table \ref{tab:ethylene_result} with the prefix 'EM' to indicate an emulator result. We study the baseline SE-QPE circuit; the cat-SE-QPE circuit (Fig.~\ref{fig:qppe_vacuum_circuit}); a mixed cat-SE-QPE circuit in which the first time evolution is performed by a controlled unitary with all subsequent performed by CSWAP-gadgets and a controlled-$U$ canonical QPE circuit. In all SE-QPE variants the reference register (in addition to the fan-out register $F$ for 'cat' variants) is measured at the end of the circuit. The reference and fan-out registers ideally return to $\ket{\overline 0}$ after each CSWAP-gadget application. Deviations from the all-zero outcome therefore indicate errors that cause leakage out of the reference subspace (including gate errors and SPAM errors on the reference qubits). In MR-cat-SE-QPE these registers are also measured mid-circuit. In Fig.~\ref{fig:abs_rel} we demonstrate that discarding any shots for which at least one of these measurements returns non-zero increases the proportion of accurate retained results.

\begin{figure}[ht]
    \centering
    \includegraphics[width=0.48\textwidth]{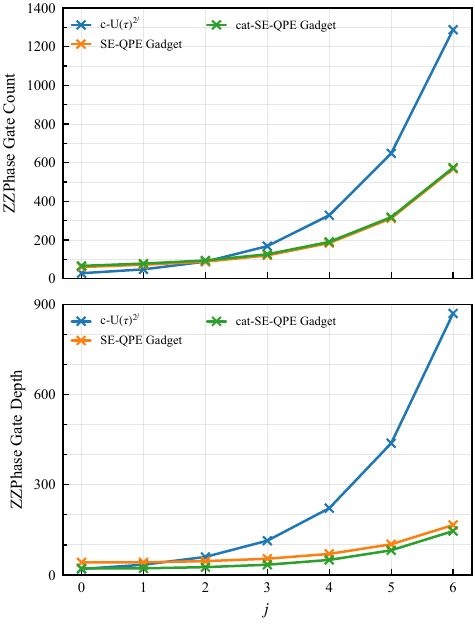}
    \caption{The 2-qubit gate count (top) and depth (bottom), with increasing $j$, per controlled-$U(\tau)^{2^j}$ and CSWAP-gadgets for SE-QPE and cat-SE-QPE. The circuits from which these metrics were derived have been optimized and compiled for the Quantinuum H2 device \cite{H2-2}. In the $j = 0$ case time evolution within each CSWAP-gadget was confined to a single register, whereas for $j>0$ the gates for time evolution were split evenly between two parallel registers.}
    \label{fig:hybrid_counts}
\end{figure}

Comparisons of the 2-qubit gate counts and depths as $j$ increases for an optimized controlled-$U(\tau)^{2^j}$ with the CSWAP-gadgets for both SE-QPE and cat-SE-QPE are presented in Fig.~\ref{fig:hybrid_counts}. The Pariser--Parr--Pople Hamiltonian for ethylene (Sec.~\ref{sec:experiments}) was used in each case. These comparisons demonstrate the increased resource gains from a ${S}_{U^{2^j}}$ (CSWAP-gadget) substitution with increasing $j$. For this system time evolution beyond the $j = 2$ case is cheaper to implement using either of the CSWAP-gadget variations than controlled-$U(\tau)^{2^j}$ in terms of 2-qubit gate count and depth. However, for $j = 0$ and $j = 1$ controlled-$U(\tau)^{2^j}$ has a lower 2-qubit gate count than its SE-QPE and cat-SE-QPE counterparts and lower depth than SE-QPE. This supports the potential benefits of mixed circuits: combining time evolution layers of controlled-$U(\tau)^{2^j}$ and CSWAP-gadgets where each is most appropriate, and motivates the inclusion of one such 'Mixed cat-SE-QPE' in the noise-modeled simulation experiments. Furthermore, comparison of the upper and lower plots in Fig.~\ref{fig:hybrid_counts} affirms the resource trade-off achieved by fan-out cat state preparation, a small increase in 2-qubit gate count for a larger decrease in 2-qubit gate depth.

\emph{Quantinuum hardware experiments.}
We execute the ethylene cat-SE-QPE, MR-cat-SE-QPE and Mixed-SE-QPE circuits on Quantinuum trapped-ion System Model H2
hardware \cite{H2-2}. Results are summarized in Table \ref{tab:ethylene_result} with the prefix 'HW' to indicate a hardware result. The cat-SE-QPE and MR-cat-SE-QPE experiments use $N=4$ system qubits, a phase register
of either $M=5$, $M=6$ or $M=7$ qubits, and  additional $N-1=3$ fan-out qubits. $\tau=10$ for $M=5$ circuits and $\tau=8$ for $M=6$ and $M=7$ circuits. The Mixed-SE-QPE experiment uses $N=4$ system qubits per register and a phase register
of $M=8$ and $\tau=8$.
The total qubit requirements are therefore
$Q_{\mathrm{cat\text{-}SE-QPE}} = 2N + M + (N-1) = 16$ when $M=5$; $Q_{\mathrm{cat\text{-}SE-QPE}} = 2N + M + (N-1) = 17$ when $M=6$; $Q_{\mathrm{cat\text{-}SE-QPE}} = 2N + M + (N-1) = 18$ when $M = 7$ and $Q_{\mathrm{SE-QPE}} = 2N + M = 16$ when $M = 8$.

We compile these circuits to the device gate set using
the compiler for H2-2 at optimization level 2 with barriers retained between each CSWAP-gadget and run
5000 shots per $M=5$, $M=6$ and $M=7$ circuit instance (only 500 shots were run for the $M=8$ Mixed-SE-QPE circuit) with dynamical decoupling threshold times of $0.03$s on the Quantinuum H2-2 quantum computer \cite{H2-2}. Removal of the barriers and/or increasing the optimization level can cause compilation with gate optimization across CSWAP-gadgets resulting in disorderly circuits with mid-circuit measurements and qubit reset gates displaced from their positions after each CSWAP-gadget.

For each dataset, after
filtering, we extract the most probable phase bitstring $x^*$, compute
$\tilde{\varphi}=x^*/2^M$, and convert to an energy estimate via
\begin{equation}
\widetilde{E}^{(b)}=-\frac{2\pi}{\tau}\left(\tilde{\varphi}+b\right),
\label{eq:energy_estimate}
\end{equation}
where $\tau$ is chosen such that the integer branch $b=0$ returns the correct energy. 
\begin{table*}[t!]
\caption{
Ethylene PPP energy estimates from emulation and hardware. The exact ground-state energy, calculated by exact diagonalization, is $-0.234282$ Ha. Statevector QPE simulation with 5, 6, 7 and 8 phase bits (see Fig.~\ref{fig:statevector}) result in ground-state energy estimations of $-0.235619$ Ha, $-0.233165$ Ha, $-0.239301$ Ha and $-0.239301$ Ha, respectively. The Phase Bits and ED Bits columns present the number of bits to store the phase estimation result and the number of 'Error Detection Bits' (i.e., bits storing measurements of fan-out or reference register qubits), respectively. Peak percentages correspond to the most common phase bitstring result in each experiment and $\widetilde{E}_{GS}$ is the energy in Ha reconstructed from this result, Eq. \eqref{eq:energy_estimate}. Filtering only retains shots for which all reference and/or fan-out register measurements return zero. This filtering is not possible for a standard implementation of canonical QPE.} 
% M7: Estimated Eigenvalue (from trot unitary) : -0.23930100291016002
% M8: Estimated Eigenvalue (from trot unitary) : -0.23930100291016002

\centering
\scriptsize
\renewcommand{\arraystretch}{1.25}
\setlength{\tabcolsep}{6pt}
\begin{tabular*}{\textwidth}{@{\extracolsep{\fill}}lccccccccc}
\hline\hline
Method & $\widetilde{E}_{GS}$ (Ha) & Total & Phase & ED & Peak ($\%$) & Peak ($\%$) & Total Gate & ZZPhase & ZZPhase\\
 & & Qubits & Bits & Bits & Raw & Filtered & Count & Count & Depth \\
\hline
EM QPE & $-0.235619$ & $9$ & $5$ & $0$ & $19.42$ & N/A & $1565$ & $655$ & $464$\\
EM SE-QPE & $-0.235619$ & $13$ & $5$ & $4$ & $30.20$ & $38.19$ & $1504$ & $537$ & $263$\\
EM Mixed-SE-QPE & $-0.235619$ & $13$ & $5$ & $4$ & $31.04$ & $36.76$ & $1407$ & $503$ & $239$\\
EM cat-SE-QPE & $-0.215984$ & $16$ & $5$ & $7$ & $16.32$ & $20.41$ & $1627$ & $567$ & $166$\\
EM MR-cat-SE-QPE & $-0.235619$ & $16$ & $5$ & $35$ & $17.28$ & $18.97$ & $1690$ & $567$ & $166$\\
HW cat-SE-QPE & $-0.235619$ & $16$ & $5$ & $7$ & $31.08$ & $36.35$ & $1627$ & $567$ & $166$\\
HW MR-cat-SE-QPE & $-0.235619$ & $16$ & $5$ & $35$ & $36.52$ & $42.77$ & $1690$ & $567$ & $166$\\
EM QPE & $-0.674952$\footnote{The results of this experiment had significant noise and no clear peak.} & $10$ & $6$ & $0$ & $6.32$\footnotemark[1] & N/A & $3106$ & $1304$ & $918$\\
EM SE-QPE & $-0.245437$ & $14$ & $6$ & $4$ & $19.56$ & $25.47$ & $2378$ & $854$ & $367$\\
EM Mixed-SE-QPE & $-0.245437$ & $14$ & $6$ & $4$ & $19.76$ & $24.03$ & $2280$ & $820$ & $343$\\
EM cat-SE-QPE & $-0.233165$ & $17$ & $6$ & $7$ & $13.28$ & $19.41$ & $2524$ & $890$ & $251$\\
EM MR-cat-SE-QPE & $-0.233165$ & $17$ & $6$ & $42$ & $14.56$ & $20.77$ & $2601$ & $890$ & $251$\\
HW cat-SE-QPE & $-0.245437$ & $17$ & $6$ & $7$ & $25.18$ & $28.32$ & $2524$ & $890$ & $251$\\
HW MR-cat-SE-QPE & $-0.245437$ & $17$ & $6$ & $42$ & $21.20$ & $25.25$ & $2601$ & $890$ & $251$\\
HW cat-SE-QPE & $-0.239301$ & $18$ & $7$ & $7$ & $19.66$ & $27.36$ & $4126$ & $1470$ & $400$\\
HW MR-cat-SE-QPE & $-0.239301$ & $18$ & $7$ & $49$ & $28.84$ & $33.04$ & $4175$ & $1470$ & $400$\\
HW Mixed-SE-QPE & $-0.239301$\footnote{A total of 500 shots were run for this experiment} & $16$ & $8$ & $4$ & $17.20$\footnotemark[2] & $22.00$\footnotemark[2] & $6847$ & $2481$ & $807$\\
\hline\hline
\end{tabular*}
\label{tab:ethylene_result}
\end{table*} 
Table~\ref{tab:ethylene_result} summarizes the extracted energies, shot statistics and circuit resources for a selection of QPE and SE-QPE experiments.
In Fig.~\ref{fig:statevector} the measured peak outcomes of $M=7$ cat-SE-QPE and MR-cat-SE-QPE hardware experiments both correspond to the statevector energy estimate. The peak outcome for MR-cat-SE-QPE has a higher probability than its cat-SE-QPE counterpart, consistent with mid-circuit fan-out and reference register reset and post-selection suppressing detectable errors. Furthermore, peak outcomes from the $M=5$ versions of MR-cat-SE-QPE on H2-2 \cite{H2-2} and QPE on H2-Emulator \cite{H2-Emulator} distributions both align with the statevector ground-state energy estimation lines. These results are consistent with SE-QPE reproducing the phase-register statistics of QPE and indicate that the circuit depth reduction (together with post-selection based on error detection) can significantly reduce noise in results. Filtered $M=6$ and $M=8$ results are in Appendix~\ref{app:6_8_plot}.
Finally, we observe improved concentration of the dominant phase peak after
filtering,  as shown in Fig.~\ref{fig:abs_rel}. These results demonstrate
that SE-QPE provides inherent error flags without modifying the phase-estimation
post-processing, and that the vacuum reference register can be used both as a
phase anchor (Sec.~\ref{sec:pcpe}) and as a symmetry-based error detection check
for particle-conserving Hamiltonians.
Comparing the impact of filtering based on measurements of vacuum register qubits versus the fan-out qubits used to prepare a cat state in Fig.~\ref{fig:abs_rel} we conclude that vacuum register filtering consistently produces a more significant improvement to results.
The agreement between results from SE-QPE and Mixed-SE-QPE is consistent with the equivalence of CSWAP-gadgets and controlled-$U(\tau)^{2^j}$. 

\emph{Reference Register and Fan-Out Qubit
Resetting.}
Investigations of the MR-cat-SE-QPE protocols on the H2-2E, H2-Emulator and H2-2 hardware device have shown that mid-circuit resetting of qubits in the reference register, in the case that it begins as the state $\ket{\overline{0}}$ (and to a lesser extent ancilla qubits used for fan-out), can contribute to a successful outcome.
It must be noted that resetting the $\ket{\overline{0}}$ reference register is only suitable if the $\ket{\overline{0}}$ state is an exact eigenstate for the chosen Hamiltonian, otherwise state information encoded on the vacuum register will be lost upon the reset. 
If this is the case each complete application of a CSWAP-gadget should return this register to $\ket{\overline{0}}$ and any deviation is the result of an error.
With increasing unitary power, $j$, the circuit depth increases, therefore the number of operations this reference register must withstand also increases causing errors to build up. 
Contrastingly, resetting the fan-out qubits can occur with no limitations on Hamiltonian choice or the state of either register. Additionally the number of operations on each cat qubit does not increase with $j$, only their idle time does.

\begin{figure}[ht]
    \centering
    \includegraphics[width=0.48\textwidth]{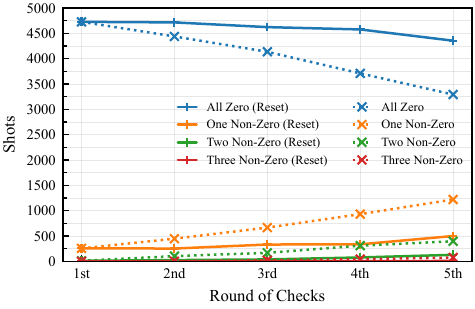}
    \caption{Comparison of bitstring results from measuring the reference register (prepared as the vacuum state) in $\tau=10$ Ha$^{-1}$ and $M=5$ cat-SE-QPE circuits, mid-circuit after each round of time evolution versus after each time evolution step without resetting the reference register qubits. All circuits were run on the Quantinuum H2-Emulator \cite{H2-Emulator}. The '1st round of checks' refers to reference register measurements after the time evolution step with $j = 0$, '2nd round' refers to after $j = 1$ and so forth. Any non-zero bit returned indicates at least one error within the circuit execution prior to that measurement.}
    \label{fig:vac_reset}
\end{figure}

Fig.~\ref{fig:vac_reset} depicts the impact of mid-circuit measurements and resetting of the reference register in 4-qubit ethylene experiments of 5000 shots on the H2-emulator for $M=5$ and $\tau=10~\mathrm{Ha}^{-1}$ \cite{H2-Emulator}. The plot shows how reference register measurements change as the number of reset layers increases. Each 'round' refers to the complete implementation of ${S}_{U^{2^j}}$ for a given $j$. The first round corresponds to $j = 0$, the second to $j = 1$ etc.,
The left plot shows the proportion of register qubits that return non-zero tends to increases with $j$ but is reduced overall when the register qubits are reset after each CSWAP-gadget.
However, the $M=6$ hardware results in Table~\ref{tab:ethylene_result} show a higher percentage of shots (both before and after filtering) result in the overall most common bitstring in the experiment using cat-SE-QPE compared to MR-cat-SE-QPE. This suggests that the additional overhead from mid-circuit measurement and reset can outweigh the benefits of resetting, so the cat-SE-QPE circuit outperforms its MR-cat-SE-QPE counterpart. %This suggests that errors associated with mid-circuit measurement and reset can overtake any suppression of errors achieved by resetting and thus the cat-SE-QPE circuit outperforms its MR-cat-SE-QPE counterpart.

\begin{figure*}[t!]
    \centering
    \includegraphics[width=0.98\textwidth]{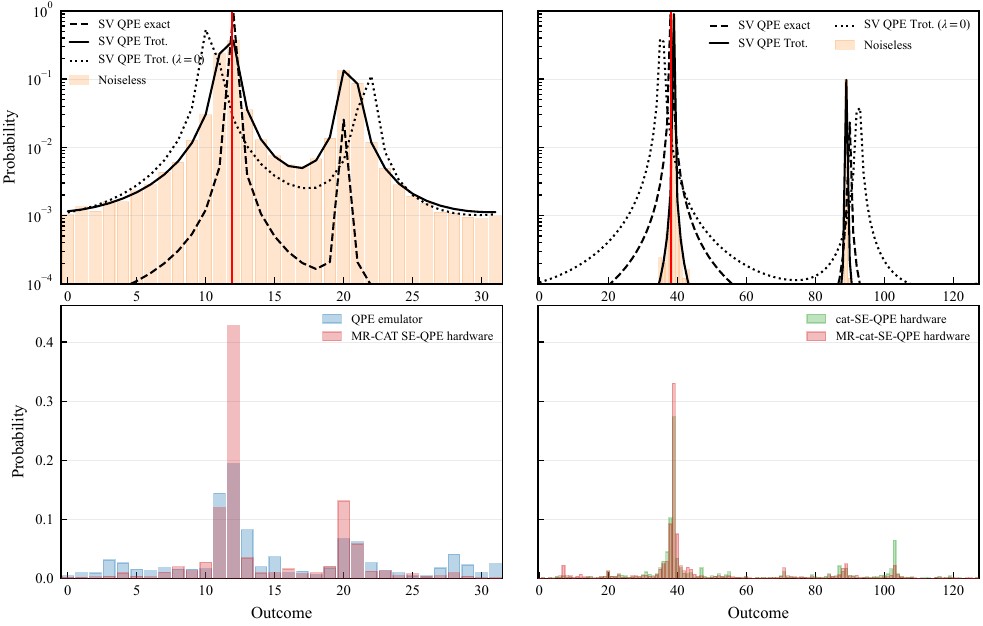}
    \caption{(Top) Noiseless SE-QPE simulations and statevector calculations for the exact evolution and Trotterization with $U(\tau) = e^{-i\tau H} \approx
e^{-i s_1 H_1}e^{-i s_2 H_2}e^{-i s_1 H_1}$ of QPE for the ethylene model Hamiltonian with a mean-field solution initial state for $\tau=10$ Ha$^{-1}$ and $M=5$ (left) and $\tau=8$ Ha$^{-1}$ and $M=7$ (right). Red lines show the exact ground-state energy estimation. The orange histograms show results from 50,000 shots of noiseless simulation of the SE-QPE circuits on the Quantinuum H2-1LE simulator \cite{H2-1LE}. (Bottom) Results from shot-based emulator and hardware experiments (all 5000 shots each) of the same Hamiltonian (described in Table \ref{tab:ethylene_result}) for $M=5$ (left) and $M=7$ (right). Distributions of filtered outcomes from MR-cat-SE-QPE circuits on the Quantinuum H2-2 device \cite{H2-2} are presented in red and only include the 3673 (left) and 4077 (right) shots that passed all error detection checks. The distribution of filtered outcomes from $M=7$ cat-SE-QPE on the Quantinuum H2-2 device \cite{H2-2} is presented in green (right) and only includes the 2716 shots that passed all end of circuit error detection checks. The outcome distribution from a QPE circuit on the Quantinuum H2-Emulator for $M=5$ \cite{H2-Emulator} is presented in blue (left).}
    \label{fig:statevector}
\end{figure*}

\section{Conclusion and outlook}
\label{sec:discussion}
We described split-evolution quantum phase estimation, a modification of phase
estimation in which controlled-$U^{2^j}$ blocks are replaced by a CSWAP-gadget acting
between a target register and a reference register, applying time
evolution without control. Using a reference eigenstate with known phase and, in particular, the vacuum for particle-conserving Hamiltonians, SE-QPE preserves the phase-register measurement statistics of canonical QPE even for initial states in a superposition of eigenstates. The two-register construction also enables parallel evolution and a native error detection check through the reference register.

We implemented circuits equivalent to $9$, $10$, $11$ and $12$ qubit canonical QPE circuits (4 state register qubits + $5$,$6$,$7$,$8$ ancilla qubits, respectively) on hardware using our split-evolution approach. The circuits were performed on the Quantinuum H2-2 device \cite{H2-2}, with at least $16$ qubits and up to $2481$ two-qubit gates, and included a natural form of error detection to inform post-selection. The 5-phase bit SE-QPE cases, on both emulator and hardware, show a similar or better probability distribution as the canonical QPE on the Quantinuum H2-Emulator \cite{H2-Emulator} around the most likely bitstring. The 6-phase bit SE-QPE cases successfully resolved the correct ground-state energy peak, whereas no clear peak could be determined from the canonical QPE equivalent on the Quantinuum H2-Emulator \cite{H2-Emulator}.

For the time evolution blocks of QPE, the leading asymptotic gains in the double-factorized chemistry model are $g_{\mathrm{CX}}^{\mathrm{count}}\simeq 2/3$, $g_{\mathrm{CX}}^{\mathrm{depth}}\simeq 3/N$, $g_{R_z}^{\mathrm{count}}\simeq 3/4$, and $g_{R_z}^{\mathrm{depth}}\simeq 3/8$, with the same scaling carrying over to $T$ metrics under a constant per-rotation synthesis cost. More importantly in practice, we show a $j$-dependent crossover and that even low bits can benefit from substitution. Both the PPP ethylene model and compiled second-order Trotter estimates across FeMoco active spaces up to (30e,30o) demonstrate this favorable trend, including synthesis and cat-CSWAP overheads.

For the PPP ethylene model, simulations and hardware experiments on Quantinuum System Model H2 devices show phase distributions consistent with canonical QPE, and the agreement between emulation of SE-QPE and Mixed-SE-QPE demonstrates the substitution in practice. The auxiliary registers provide useful error flags and filtering increases the probability of the dominant phase peak, from both vacuum-register and cat-register checks. The split-evolution approach extends hardware chemistry phase-estimation demonstrations to a four-qubit system register with an explicit inverse QFT.

SE-QPE trades reduced depth for extra qubits and CSWAP layers, and its most favorable depth scalings rely on architecture-dependent assumptions such as all-to-all connectivity. The circuit depth associated with the CSWAP layers can be reduced by parallelization at the cost of additional fan-out qubits to prepare a cat state. Near future work is to optimize mixed QPE and SE-QPE substitutions under device constraints, exploit early exit from reference register checks, and use the method in early fault-tolerant pipelines. The same primitive also supports phase-difference and gap estimation by preparing different states on the two lanes, and may also be useful beyond QPE when phase estimation appears only as a subroutine in a larger quantum algorithm.
 
\begin{acknowledgments} \label{sec:acknowledgements}
We thank Duncan Gowland and Yuta Kikuchi for their comprehensive review of the manuscript and for their valuable suggestions and insights. We thank Robert Anderson for the discussions about factorized swap-test and compute--uncompute protocols in InQuanto. Additionally, we acknowledge the entire chemistry team for their insightful discussions and feedback throughout this work.
\end{acknowledgments}

MCR and DZM were responsible for the main development, analysis of the method and preparation of the manuscript. CG, DMR and DZM were responsible for resource estimation for FeMoco. MCR, LS and DZM were responsible for preparation of the ethylene model and hardware calculations.

% Uncomment if you have a .bib file and citations
\bibliographystyle{apsrev4-2}
\bibliography{references}

\appendix

\section{Jordan--Wigner encoded PPP Hamiltonian for ethylene in the (1,1) sector}
\label{app:ethylene_model}
For ethylene the PPP parametrization in Ref. \cite{zhang2011excitation} is constructed such that
$t_{11}=t_{22}$, $t_{12}=t_{21}$, $\gamma_{11}=\gamma_{22}$, $\gamma_{12}=\gamma_{21}$.
We introduce three effective couplings $\alpha$, $\beta_{1}$, $\beta_{2}$ as the coefficients of the qubit operators in the $N_{\uparrow}=1$, $N_{\downarrow}=1$ sector,
\begin{equation}
\begin{split}
H^{(\mathrm{C_2H_4})}_{(1,1)}
 =& C_{\mathrm{tot}}' + 
\alpha\,(X_{1}X_{2}+Y_{1}Y_{2}+X_{3}X_{4}+Y_{3}Y_{4})
\\[2pt]
&+\beta_{1}\,(Z_{1}Z_{4}+Z_{2}Z_{3}) + \beta_{2}\,(Z_{1}Z_{3}+Z_{2}Z_{4}),
\end{split}
\end{equation} and the Jordan--Wigner mapping with 
\begin{equation}
(1\uparrow,2\uparrow,1\downarrow,2\downarrow)
\;\longleftrightarrow\;
(1,2,3,4).
\end{equation}
Further reductive transformations are possible based on symmetries, but for the experiment this Hamiltonian is not simplified further. Matching this to the PPP Hamiltonian in Jordan--Wigner form gives the
relations between the qubit couplings and the original one- and
two-electron parameters,
$t_{12}=t_{21}=2\alpha$, $\gamma_{12}=\gamma_{21}=4\beta_{1}$, $\gamma_{11}=\gamma_{22}=4\beta_{2}$. In particular for ethylene, 
$C_{\mathrm{tot}}' = -0.347936~\text{Ha}$,
$\alpha = -0.055557~\text{Ha}$,
$\beta_{1} = 0.067525~\text{Ha}$ and
$\beta_{2} = 0.104616~\text{Ha}$ have been used. For the purpose of phase estimation we drop the constant term, and introduce \begin{equation}
H = H_1 + H_2,
\end{equation} where 
\begin{equation}H_1 = \alpha\,(X_{1}X_{2}+Y_{1}Y_{2}+X_{3}X_{4}+Y_{3}Y_{4})\end{equation} and \begin{equation}H_2 = \beta_{1}\,(Z_{1}Z_{4}+Z_{2}Z_{3}) + \beta_{2}\,(Z_{1}Z_{3}+Z_{2}Z_{4}).\end{equation} 
The singlet ground-state can be written as
\begin{equation}
\begin{split}
\ket{\psi(\vartheta)}
&=
\frac{\cos\vartheta}{\sqrt{2}}\bigl(\ket{1010}+\ket{0101}\bigr)
+\\&+
\frac{\sin\vartheta}{\sqrt{2}}\bigl(\ket{0110}+\ket{1001}\bigr),
\end{split}
\end{equation}
where the exact ground-state is with $0<\vartheta^*<\pi/2$
\begin{equation}
\ket{S_{0}} = \ket{\psi(\vartheta^*)},\qquad
\tan(2\vartheta^*)
=
\frac{2\alpha}{\beta_{2}-\beta_{1}},
\end{equation} that is $\vartheta^* = 0.94648805$. The mean-field solution is taken as $\ket{S_0^{\mathrm{MF}}} = \ket{\psi(\vartheta_{\mathrm{MF}})}$ where $\vartheta_{\mathrm{MF}} = \frac{\pi}{4}$ and $\braket{S_0}{S_0^{\mathrm{MF}}} = 0.987053$.
The triplet state in this sector is fixed, 
\begin{equation}
\ket{T_{0}} = \frac{1}{\sqrt{2}}\bigl(\ket{1001}-\ket{0110}\bigr)
\end{equation} and the vacuum reference state
\begin{equation}
\ket{\overline{0}} = \ket{0000}.
\end{equation}
% Overlap   <S0|S0_MF>    = +0.987053+0.000000j
% Fidelity |<S0|S0_MF>|^2 = +0.974274
% Energy S0      = -0.234282   (+ctotp = -0.582217)
% Energy S0 MF   = -0.222228   (+ctotp = -0.570163)
% Energy corr S0 = +0.012054
% Energy T0      = -0.074182   (+ctotp = -0.422117)
% Energy Vacuum  = +0.344282   (+ctotp = -0.003653)
% tau = +8.000000
% M = +6.000000
% Lead Trot Error = +0.026089
% QPE Error       = +0.006136
% Total Error     = +0.032225
% After eliminating the lead trot error with lam=0.000917161520666667:
% s1       = +4.469587
% s2       = +8.000000

The energies are 
$E_{S_{0}} = -0.234282~\text{Ha}$, 
$E_{S_{0}^{\mathrm{MF}}} = -0.222228~\text{Ha}$, 
$E_{T_{0}} \approx -0.074182~\text{Ha}$, 
$E_{\overline{0}} \approx +0.344282~\text{Ha}$, 
$\Delta_{ST} = E_{T_{0}}-E_{S_{0}} \approx 0.160100~\text{Ha}$.

\begin{figure}[t]
\centering
\begin{quantikz}
\lstick{$\ket{0}_{1}$} & \gate[1]{\text{$H$}} & \ctrl{2} & \ctrl{1} & \qw & \qw \\
\lstick{$\ket{0}_{2}$} & \qw & \qw & \targ{} & \gate[1]{\text{$X$}} & \qw \\
\lstick{$\ket{0}_{3}$} & \gate[1]{\text{$R_Y(2\vartheta/\pi)$}} & \targ{} & \ctrl{1} & \qw & \qw \\
\lstick{$\ket{0}_{4}$} & \qw & \qw & \targ{} & \gate[1]{\text{$X$}} & \qw \\
\end{quantikz}
\caption{Ansatz circuit $\ket{\psi(\vartheta)}$.}
\label{fig:ansatz_state}
\end{figure}
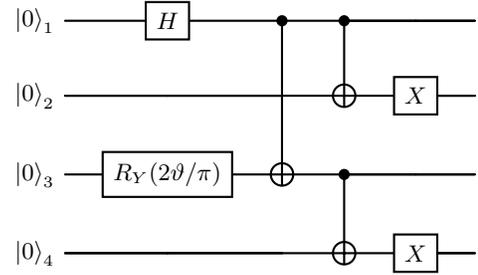

\section{Trotter approximation of the PPP Hamiltonian for ethylene}
\label{app:trotter_ethylene}
We approximate the QPE time evolution unitary $U(\tau)=e^{-i\tau H}$ by the symmetric Strang-like product
\begin{equation}
U_{\mathrm{trot}}(s_1,s_2)
=
e^{-i s_1 H_1}e^{-i s_2 H_2}e^{-i s_1 H_1}.
\end{equation}
The values of $s_1, s_2$ can be determined from Trotter product formulae with perturbation methods \cite{Poulin2015TrotterStep}. In this paper we obtain these values to minimize the error on the ground-state energy and the reference state energies. Since $\ket{\overline{0}}=\ket{0000}$ and $\ket{T_0}$ satisfy $H_1\ket{\overline{0}}=0$ and $H_1\ket{T_0}=0$,
their phases under $U_{\mathrm{trot}}$ are controlled entirely by $H_2$.
Thus we set
\begin{equation}
s_2=\tau,
\end{equation}
so that the vacuum and triplet phases are reproduced exactly at evolution time $\tau$, regardless of $s_1$. For that we set
\begin{equation}
s_1=\frac{\tau}{2}+\lambda \tau^3,
\end{equation}
where we need to find a $\lambda$ so that the product formula stays a second-order approximation of $e^{-i\tau H}$. Using the Baker--Campbell--Hausdorff expansion for the symmetric product \cite{Poulin2015TrotterStep, blunt2025montecarloapproachbound}, one can write
\begin{equation}
U_{\mathrm{trot}}(\tau)
=
\exp\!\left[-i\tau\left(H+\tau^2\,V^{(2)}+O(\tau^4)\right)\right],
\end{equation}
with a correction 
\begin{equation}
V^{(2)}
=
2\lambda H_1
+\Delta H^{(2)},
\end{equation}
where the Strang-splitting term is
\begin{equation}
\Delta H^{(2)}
=
\frac{1}{24}\left[H_1,\left[H_1,H_2\right]\right]
-\frac{1}{12}\left[H_2,\left[H_2,H_1\right]\right].
\end{equation}
Therefore to leading order, the eigenvalue returned by QPE on $U_{\mathrm{trot}}(\tau)$ corresponds to the ground-state energy of the
effective Hamiltonian $H_{\mathrm{eff}}=H+\tau^2 V^{(2)}$ \cite{Poulin2015TrotterStep}. If $\ket{E_0}$ is the exact ground-state of $H$ with $H\ket{E_0}=E_0\ket{E_0}$, then first-order perturbation theory results in
\begin{equation}
E_0^{(\mathrm{trot})}
=
E_0
+\tau^2\,\bra{E_0}V^{(2)}\ket{E_0}
+O(\tau^4).
\end{equation}
Therefore a simple way to minimize the error bias is to set
$\bra{E_0}V^{(2)}\ket{E_0}=0$, which yields
\begin{equation}
\lambda
=
-\frac{1}{2}\,
\frac{\bra{E_0}\Delta H^{(2)}\ket{E_0}}{\bra{E_0}H_1\ket{E_0}},
\end{equation} and the leading error on the energy is improved from $O(\tau^2)$ to $O(\tau^4)$. Typically $\ket{E_0}$ is not known exactly, however if an approximate state $\ket{\psi}$ is available with significant overlap with the ground-state, one can attempt to use \begin{equation}
\lambda
\approx
-\frac{1}{2}\,
\frac{\bra{\psi}\Delta H^{(2)}\ket{\psi}}
{\bra{\psi}H_1\ket{\psi}},
\end{equation} to reduce the ground-state energy Trotter bias in QPE. In particular, if state $\ket{S_0^{\mathrm{MF}}}$ is used, we find $\lambda \approx 0.00091716~\text{Ha}^2$. This modification enables the choice of larger $\tau$, for the same accuracy, which is beneficial to an economical QPE.

\section{Choice of the QPE time step $\tau$}
\label{app:timestep}
For an energy eigenstate $H\ket{E}=E\ket{E}$, the ideal phase relation is
\begin{equation}
U(\tau)\ket{E}=e^{-i\tau E}\ket{E}=e^{i2\pi \varphi}\ket{E},
\end{equation} where $\varphi=\left(-\frac{E\tau}{2\pi}\right)\bmod 1$ and $\varphi\in[0,1)$. With $M$ phase bits, QPE returns an $M$-bit fraction $\tilde{\varphi}=y/2^M$, ($y$ being the bitstring value of the QPE outcome). Because $\varphi$ is defined mod $1$, the energy inversion is only defined up to an
integer branch $b\in\mathbb{Z}$ such that
\begin{equation}
-\frac{E\tau}{2\pi}=\varphi+b,
\qquad
b\in\mathbb{Z}.
\end{equation}
Given the phase estimate $\tilde{\varphi}$, the corresponding energy estimates are
\begin{equation}
\tilde{E}^{(b)}=-\frac{2\pi}{\tau}\left(\tilde{\varphi}+b\right),
\qquad
b\in\mathbb{Z}.
\end{equation}
If a $E\in[E_{\min},E_{\max}]$ is known, $b$ is chosen such that $\tilde{E}^{(b)}\in[E_{\min},E_{\max}]$, and this choice is unique when $\tau\left(E_{\max}-E_{\min}\right)<2\pi$. In practice, a reference energy from a classical approximation $E_{\mathrm{ref}}$ can be also used, by picking the
integer $b$ that makes $\tilde{E}^{(b)}$ closest to $E_{\mathrm{ref}}$, formally 
\begin{equation}
b=\left\lfloor -\frac{E_{\mathrm{ref}}\tau}{2\pi}-\tilde{\varphi}+\frac{1}{2}\right\rfloor.
\end{equation}
The phase grid spacing is $2^{-M}$, hence the energy grid spacing is
\begin{equation}
\Delta E_{\mathrm{grid}}(M,\tau)=\frac{2\pi}{2^M \tau}.
\end{equation}
A representative QPE discretization error scale is
\begin{equation}
\delta E_{\mathrm{QPE}}(M,\tau)\sim \frac{1}{2}\Delta E_{\mathrm{grid}}
=\frac{\pi}{2^M \tau}.
\label{eq:Resolution}
\end{equation}
Thus, increasing $\tau$ reduces the QPE discretization error but increases the Trotter error. Assuming the Trotter error is small for the ground-state energy after the optimization, with the choice of $\tau = 8~\text{Ha}^{-1}$ and $M=6$, $M=7$ or $M=8$ the discretization error dominates. That is $s_1 = +4.469587~\text{Ha}^{-1}$, $s_2 = 8~\text{Ha}^{-1}$ and for $M=6$ $\delta E_{\mathrm{QPE}}(M,\tau)=6.135923$ mHa; for $M=7$ $\delta E_{\mathrm{QPE}}(M,\tau)=3.067962$ mHa and for $M=8$ $\delta E_{\mathrm{QPE}}(M,\tau)=1.533981$ mHa. Choosing $\tau = 10~\text{Ha}^{-1}$ and $M=5$ means discretization error is further from chemical precision with $s_1 = +5.917162~\text{Ha}^{-1}$, $s_2 = 10~\text{Ha}^{-1}$ and $\delta E_{\mathrm{QPE}}(M,\tau)=9.817477$ mHa.

\section{Trotter formula and gate counts}
For a resource counter for gate $G$, $M_G \in \{C_G, D_G\}$,
\begin{equation*}
M_G\!\left[U^{(1)}\right]
=
(L+2)M_G[W] + M_G[U_0] + L\,M_G[U_l]
\label{eq:app_M_firstorder}
\end{equation*}
and 
\begin{equation*}
M_G\!\left[U^{(2)}\right]
=
2(L+1)M_G[W] + 2M_G[U_0] + (2L-1)M_G[U_l].
\label{eq:app_M_secondorder}
\end{equation*}

For controlled steps, basis rotations remain uncontrolled, and only diagonal kernels are replaced by their controlled versions. Therefore the metrics above hold with $U_0 \mapsto \mathrm{c}U_0$ and
$U_l \mapsto \mathrm{c}U_l$.

\begin{figure*}[t]
  \centering
  \includegraphics{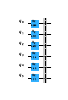}
  \caption{Diagonal one-body primitive $U_0$ for $N=6$, with explicit layer barriers. Analytical formulas (for even N) give $C_{\mathrm{CX}}=0$, $C_{R_z}=N=6$, $D_{\mathrm{CX}}=0$, $D_{R_z}=1$. Compiled circuit (without barriers) gives $C_{\mathrm{CX}}=0$, $C_{R_z}=6$, $C_{H}=0$, $D_{\mathrm{CX}}=0$, $D_{R_z}=1$, $D_{H}=0$.}
  \label{fig:primitive_u0_n6}
\end{figure*}

\begin{figure*}[t]
  \centering
  \includegraphics{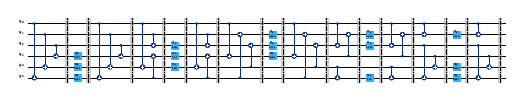}
  \caption{Diagonal two-body primitive $U_l$ for $N=6$, with explicit layer barriers. Analytical formulas (for even N) give $C_{\mathrm{CX}}=N(N-1)=30$, $C_{R_z}=\frac{N(N-1)}{2}=15$, $D_{\mathrm{CX}}=2(N-1)=10$, $D_{R_z}=N-1=5$. Compiled circuit (without barriers) gives $C_{\mathrm{CX}}=30$, $C_{R_z}=15$, $C_{H}=0$, $D_{\mathrm{CX}}=10$, $D_{R_z}=5$, $D_{H}=0$.}
  \label{fig:primitive_ul_n6}
\end{figure*}

\begin{figure*}[t]
  \centering
  \includegraphics{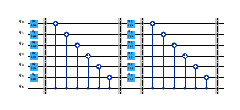}
  \caption{Controlled one-body primitive $\mathrm{c}U_0$ for $N=6$, with explicit layer barriers. Analytical formulas (for even N) give $C_{\mathrm{CX}}=2N=12$, $C_{R_z}=2N=12$, $D_{\mathrm{CX}}=2N=12$, $D_{R_z}=2$. Compiled circuit (without barriers) gives $C_{\mathrm{CX}}=12$, $C_{R_z}=12$, $C_{H}=0$, $D_{\mathrm{CX}}=12$, $D_{R_z}=2$, $D_{H}=0$.}
  \label{fig:primitive_cu0_n6}
\end{figure*}

\begin{figure*}[t]
  \centering
  \includegraphics{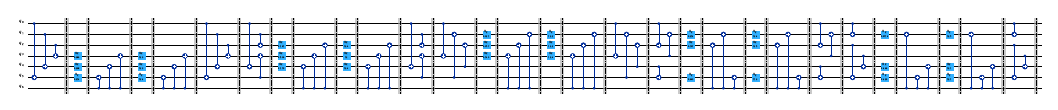}
  \caption{Controlled two-body primitive $\mathrm{c}U_l$ for $N=6$, with explicit layer barriers. Analytical formulas (for even N) give $C_{\mathrm{CX}}=2N(N-1)=60$, $C_{R_z}=N(N-1)=30$, $D_{\mathrm{CX}}=(N+2)(N-1)=40$, $D_{R_z}=2(N-1)=10$. Compiled circuit (without barriers) gives $C_{\mathrm{CX}}=60$, $C_{R_z}=30$, $C_{H}=0$, $D_{\mathrm{CX}}=36$, $D_{R_z}=10$, $D_{H}=0$.}
  \label{fig:primitive_cul_n6}
\end{figure*}

\begin{figure*}[t]
  \centering
  \includegraphics{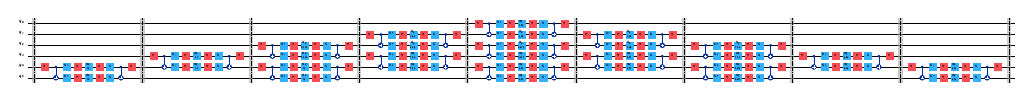}
  \caption{Basis rotation $W$ for $N=6$, with explicit layer barriers. Analytical formulas (for even N) give $C_{\mathrm{CX}}=N(N-1)=30$, $C_{R_z}=N(N-1)=30$, $D_{\mathrm{CX}}=2(2N-3)=4N-6=18$, $D_{R_z}=2N-3=9$. Compiled circuit (without barriers) gives $C_{\mathrm{CX}}=30$, $C_{R_z}=30$, $C_{H}=90$, $D_{\mathrm{CX}}=18$, $D_{R_z}=9$, $D_{H}=28$.}
  \label{fig:primitive_w_full_n6}
\end{figure*}

\begin{figure*}[t]
  \centering
  \includegraphics{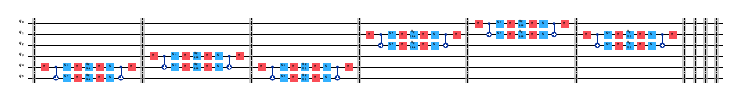}
  \caption{Spin-block basis rotation $W = W_{\uparrow}  \oplus W_{\downarrow} $ for $N=6$, with explicit layer barriers. Analytical formulas (for even N) give $C_{\mathrm{CX}}=\frac{N(N-2)}{2}=12$, $C_{R_z}=\frac{N(N-2)}{2}=12$, $D_{\mathrm{CX}}=2N-6=6$, $D_{R_z}=N-3=3$. Compiled circuit (without barriers) gives $C_{\mathrm{CX}}=12$, $C_{R_z}=12$, $C_{H}=36$, $D_{\mathrm{CX}}=6$, $D_{R_z}=3$, $D_{H}=10$.}
  \label{fig:primitive_w_block_n6}
\end{figure*}

Figs.~\ref{fig:primitive_u0_n6}, \ref{fig:primitive_ul_n6}, \ref{fig:primitive_cu0_n6}, \ref{fig:primitive_cul_n6}, \ref{fig:primitive_w_full_n6}, and \ref{fig:primitive_w_block_n6} show the primitive circuits used in the resource model. If we exploit sparsity of $W$, as in Fig.~\ref{fig:primitive_w_block_n6}, the gain factors reduce to 
\begin{equation}
g_{\mathrm{CX}}^{\mathrm{count}} \simeq \frac{3}{5},
\qquad
g_{\mathrm{CX}}^{\mathrm{depth}} \simeq \frac{2}{N}
\end{equation}
and
\begin{equation}
g_{R_z}^{\mathrm{count}} \simeq \frac{2}{3},
\qquad
g_{R_z}^{\mathrm{depth}} \simeq \frac{1}{3}.
\end{equation}

\begin{figure*}[t]
  \centering
  \includegraphics[width=\textwidth]{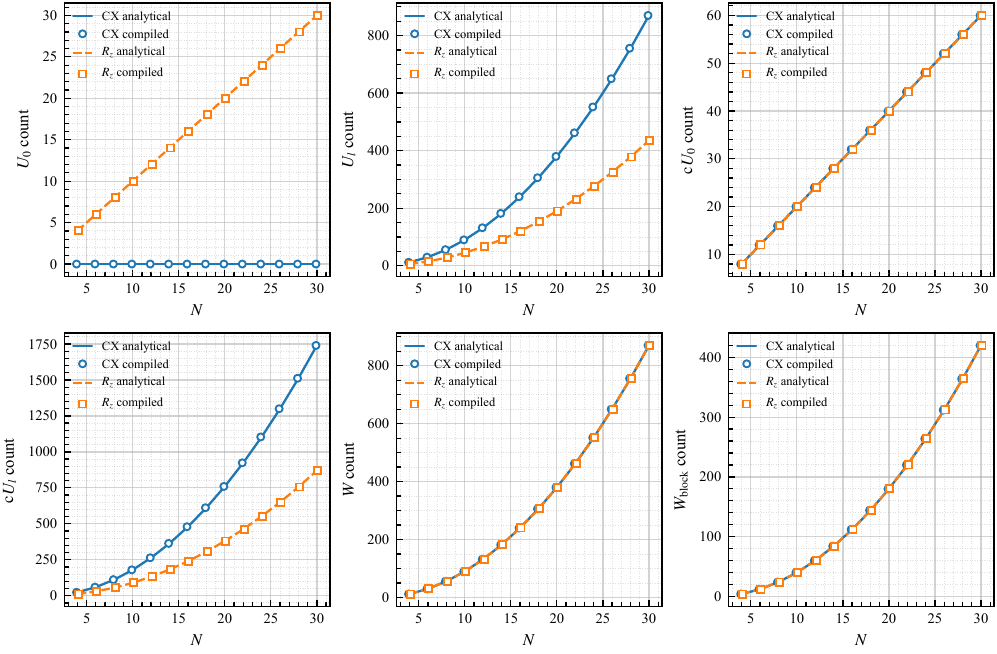}
  \caption{Resource metrics for circuit building blocks. Panels show $U_0$, $U_l$, $\mathrm{c}U_0$, $\mathrm{c}U_l$, $W$, and $W_{\mathrm{block}}$ (spin-blocked version of W). Blue denotes CX and orange denotes $R_z$. Solid/dashed lines are analytical formulas, hollow markers are compiled values from barrier-free circuits.}
  \label{fig:si_primitive_pred_comp_counts}
\end{figure*}

\begin{figure*}[t]
  \centering
  \includegraphics[width=\textwidth]{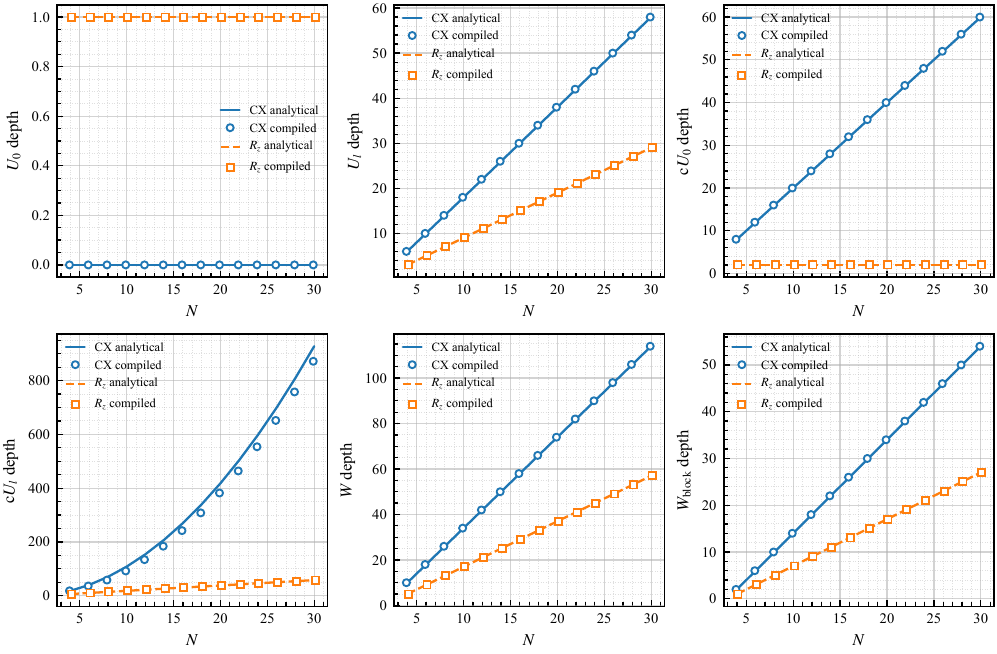}
  \caption{Resource metrics for circuit building blocks. Panels show $U_0$, $U_l$, $\mathrm{c}U_0$, $\mathrm{c}U_l$, $W$, and $W_{\mathrm{block}}$ (spin-blocked version of W). Blue denotes CX and orange denotes $R_z$. Solid/dashed lines are analytical formulas, hollow markers are compiled values from barrier-free circuits.}
  \label{fig:si_primitive_pred_comp_depths}
\end{figure*}

\section{Noiseless simulations of QPE, SE-QPE and CU-QPE}
\label{app:noiselessCU}

\begin{figure}[ht]
    \centering
    \includegraphics[width=0.48\textwidth]{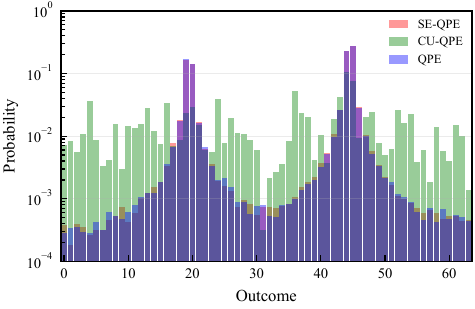}
    \caption{Results from 50,000 shot noiseless simulations of canonical QPE (blue), CU-QPE (green) and SE-QPE (red) circuits on the Quantinuum H2-1LE simulator \cite{H2-1LE}. In each case $\tau=8$ Ha$^{-1}$ and $M=6$.}
    \label{fig:noiselessCUplot}
\end{figure}

We use noiseless simulations to demonstrate the inequivalence between compute--uncompute quantum phase estimation (CU-QPE) and canonical QPE when $\ket{\psi}$ is not an eigenstate of $U(\tau)$. CU-QPE is a canonical QPE circuit in which each controlled time evolution step is replaced by a $S_{U^{2^j}}^{\mathrm{CU}}$ gadget. In Fig. \ref{fig:noiselessCUplot} we plot the outcomes from 50,000 shots each of SE-QPE, CU-QPE and QPE on the Quantinuum H2-1LE simulator \cite{H2-1LE} for the ethylene PPP Hamiltonian with $\tau=8$ Ha$^{-1}$ and $M=6$. For this investigation the ansatz circuit $\ket{\psi(\vartheta)}$ (Fig. \ref{fig:ansatz_state}) was perturbed from the exact ground-state of the ethylene PPP Hamiltonian by choosing $\vartheta=0.2$.
The overall distribution of results from CU-QPE is noticeably different to QPE and SE-QPE. It has a much broader spread and lacks definition at the two clear peaks displayed by both QPE and SE-QPE. This simulation result also reaffirms the equivalence between SE-QPE and QPE (demonstrated in Fig. \ref{fig:statevector}) for a non-eigenstate $\ket{\psi}$, since their distributions remain indistinguishable, within sampling uncertainty, even for initial states without a single dominant eigenstate in the superposition.

\section{Circuit compiled for hardware}
\label{app:hardware_circuits}

\begin{figure*}[t]
  \centering
  \includegraphics{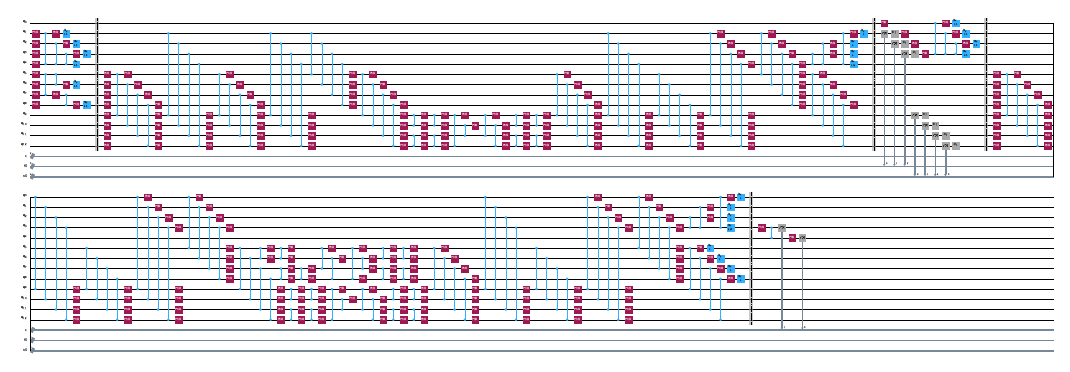}
  \caption{A MR-cat-SE-QPE circuit to determine 2 digits of the ethylene PPP Hamiltonian ground-state energy phase bitstring, compiled for the Quantinuum System Model H2 devices. Purple boxes denote PhasedX gates, blue lines indicate ZZPhase gates, and blue boxes denote Rz gates.}
  \label{fig:cat_MR-SE-QPE_comp_circuit}
\end{figure*}

To illustrate the SE-QPE algorithm in practice we compile a 2 bit example of the cat-MR-SE-QPE circuit for the ethylene PPP Hamiltonian with the ansatz circuit $\ket{\psi(\vartheta=\frac{\pi}{4})}$ (see Fig. \ref{fig:ansatz_state}), for the Quantinuum H2-2 device \cite{H2-2}. This circuit is presented in Fig. \ref{fig:cat_MR-SE-QPE_comp_circuit}.

\section{Six and eight bit statevector and hardware results}
\label{app:6_8_plot}

\begin{figure*}[p]
    \centering
    \includegraphics[width=0.98\textwidth]{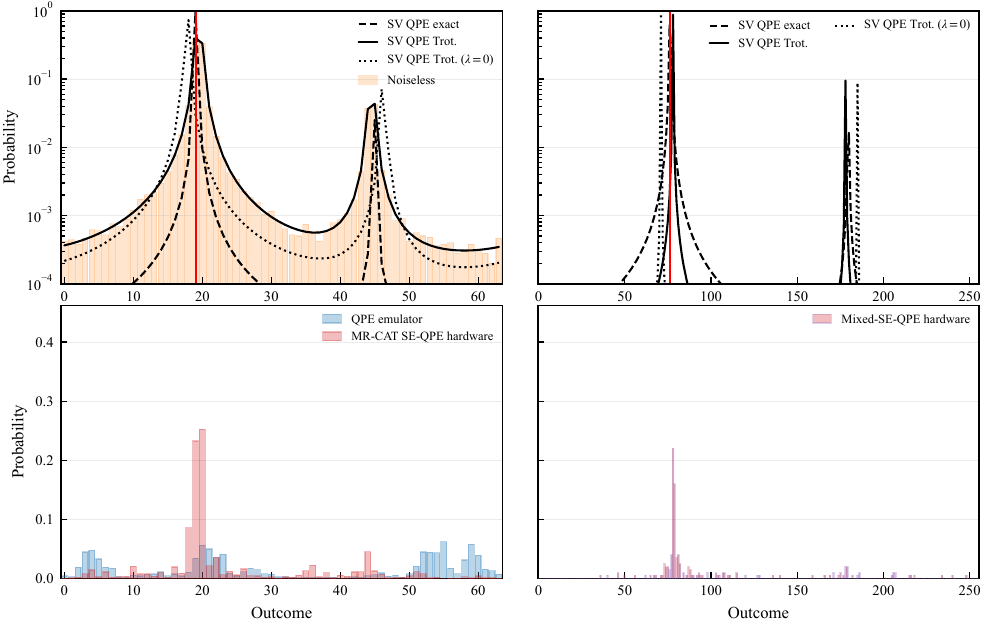}
    \caption{(Top) Statevector calculations for the exact evolution and Trotterization with $U(\tau) = e^{-i\tau H} \approx
e^{-i s_1 H_1}e^{-i s_2 H_2}e^{-i s_1 H_1}$ of QPE for the ethylene model Hamiltonian with a mean-field solution initial state for $\tau=8$ Ha$^{-1}$ with $M=6$ (left) and $M=8$ (right). Red lines show the exact ground-state energy estimation. The orange histogram (left) shows results from 50,000 shots of noiseless simulation of the $M=6$ SE-QPE circuit on the Quantinuum H2-1LE simulator \cite{H2-1LE}. (Bottom) Results from shot-based emulator and hardware experiments of the same Hamiltonian (described in Table \ref{tab:ethylene_result}) for $M=6$ (left) and $M=8$ (right). The distributions of filtered outcomes from 5000 shots of the $M=6$ MR-cat-SE-QPE circuit and 500 shots of the $M=8$ Mixed-SE-QPE on the Quantinuum H2-2 device \cite{H2-2} are presented in red and purple respectively. These distributions only include the 3161 (left) and 200 (right) shots that passed all error detection checks. The outcome distribution from a QPE circuit on the Quantinuum H2-Emulator for $M=6$ \cite{H2-Emulator} is presented in blue (left).}
\label{fig:statevector_m6m8}
\end{figure*}

 In Fig.~\ref{fig:statevector_m6m8} we plot the filtered MR-cat-SE-QPE and Mixed-SE-QPE hardware results described in Table \ref{tab:ethylene_result} for $\tau=8$ with $M=6$ and $M=8$ respectively. The peaks from both of these distributions align with peaks in their corresponding Trotterized statevector calculations (plotted as solid black lines in the top panel of Fig.~\ref{fig:statevector_m6m8}). Unlike the $M=6$ MR-cat-SE-QPE, no clear peak could be determined from the results of an $M=6$ QPE circuit on the Quantinuum H2-Emulator \cite{H2-Emulator} (as shown in blue on the bottom left panel of Fig.~\ref{fig:statevector_m6m8}).

\section{FeMoco active-space scaling inputs}
\label{app:femoco_l_scaling}

\begin{figure*}[t]
    \centering
    \includegraphics[height=0.58\textheight,keepaspectratio]{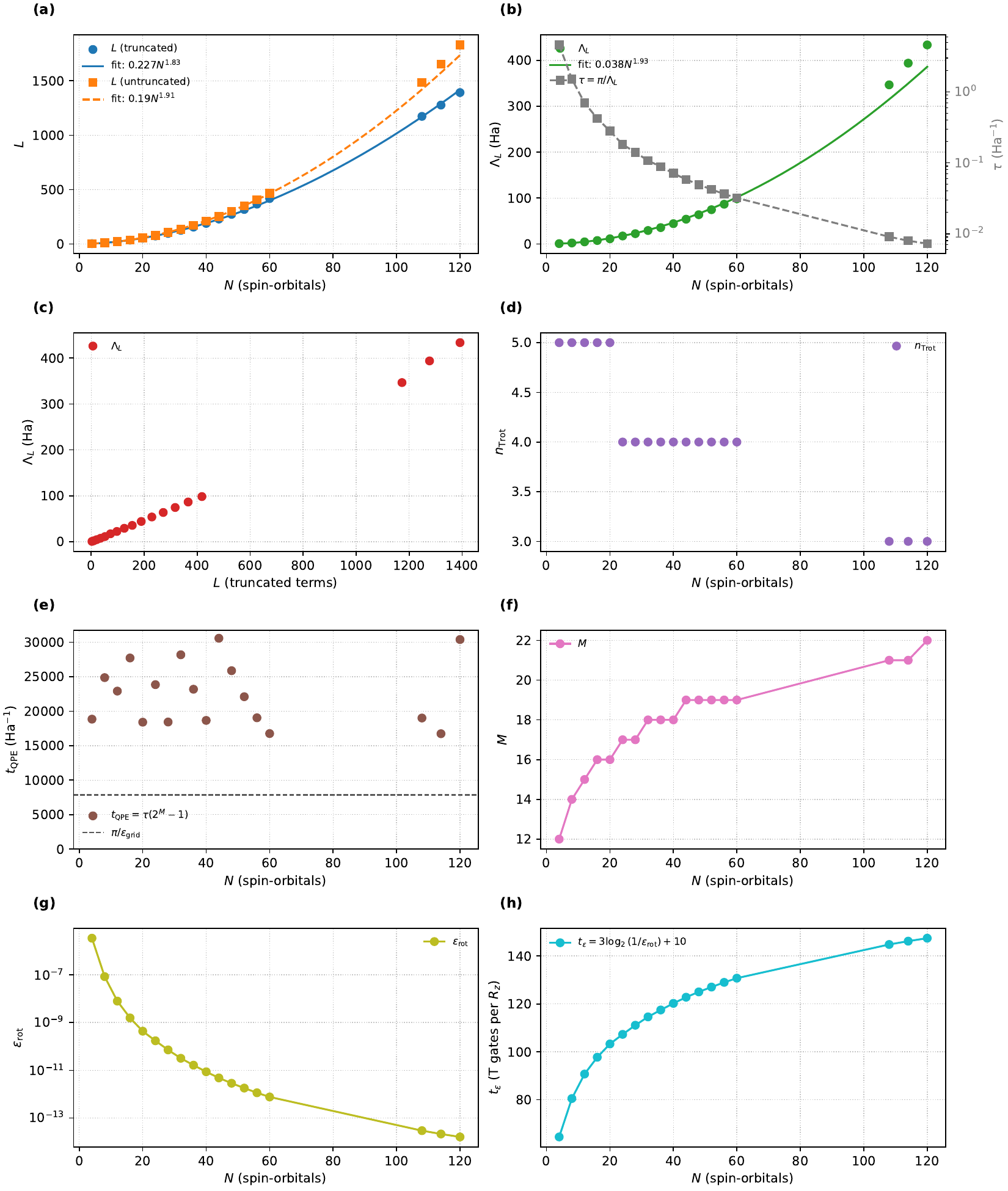}
    \caption{FeMoco active-space scaling data and derived phase-estimation parameters used in the resource estimates. (a) Retained number of double-factorized terms $L$ after truncation, together with the untruncated term count. Solid and dashed curves are empirical power-law fits over the shown active spaces. (b) Retained coefficient-norm scale $\Lambda_L$ and the corresponding QPE evolution time $\tau=\pi/\Lambda_L$. (c) Relation between $\Lambda_L$ and retained $L$. (d) Number of second-order Trotter steps from the heuristic in Sec.~\ref{sec:femoco_resources}. (e) Total accumulated QPE evolution time $t_{\mathrm{QPE}}=\tau(2^M-1)$, with $\pi/\varepsilon_{\mathrm{grid}}$ shown as a reference scale. (f) Phase-register size $M$ used in the resource estimates. (g) Per-rotation synthesis precision $\varepsilon_{\mathrm{rot}}$. (h) Corresponding $T$-gate synthesis estimate $t_\epsilon=3\log_2\left(1/\varepsilon_{\mathrm{rot}}\right)+10$ per $R_z$ rotation.}
    \label{fig:femoco_l_norm_scaling}
\end{figure*}

Figure~\ref{fig:femoco_l_norm_scaling} shows the intermediate active-space quantities used to construct the FeMoco resource estimates in Sec.~\ref{sec:femoco_resources}. The figure details data obtained from the Cholesky double-factorized Hamiltonians from quantities derived by the resource model. For each AS, the retained double-factorized expansion is chosen by requiring the discarded coefficient-norm bound to be below $\varepsilon_{\mathrm{trunc}}$. This determines the retained number of factors $L$ and the coefficient-norm scale $\Lambda_L$ in Eq.~\eqref{eq:femoco_LambdaH}. The evolution time for the unitary is set by
\begin{equation}
\tau=\frac{\pi}{\Lambda_L},
\end{equation} which is a conservative choice because $\Lambda_L$ bounds the spectral radius, giving
$\tau(E_{\max}-E_{\min})\le 2\pi$.
The growth of $\Lambda_L$ with active-space size therefore shortens the base evolution time. The derived quantities have a stepwise structure with size because for multiple reasons, such as Trotterization and the phase grid choice. The number of phase bits, $M$ changes only when the phase grid requires another bit, while total phase estimation time $t_{\mathrm{QPE}}=\tau(2^M-1)$ decreases between such jumps because $\tau$ decreases.

The Trotter and synthesis panels use the same error allocation as Sec.~\ref{sec:femoco_resources}. In particular,
\begin{equation}
n_{\mathrm{Trot}}
=
\left\lceil
c_{\mathrm{TS}}
\frac{\tau N^\xi}{\sqrt{\varepsilon_{\mathrm{TS}}}}
\right\rceil
\end{equation}
with $\xi=1.7$. The scaling constant $c_{\mathrm{TS}}$ is calibrated from the reference Trotter error
\begin{equation}
\delta_{\mathrm{ref}}
=
\left\|e^{-i\tau_{\mathrm{ref}} H}
-U^{(2)}(\tau_{\mathrm{ref}})\right\|
\end{equation}
computed for AS (6e,6o), with $N_{\mathrm{ref}}=12$ and $\tau_{\mathrm{ref}}=1~\mathrm{Ha}^{-1}$. Using $\delta_{\mathrm{ref}}=0.01630$ gives
\begin{equation}
c_{\mathrm{TS}}
=
\frac{\sqrt{\delta_{\mathrm{ref}}/\tau_{\mathrm{ref}}^{3}}}
{N_{\mathrm{ref}}^{\xi}}
\approx0.00187 
\end{equation} in appropriate unit
and the per-rotation synthesis precision is allocated over one simulated $U(\tau)$ as
\begin{equation}
\varepsilon_{\mathrm{rot}}
=
\frac{\tau\varepsilon_{\mathrm{synth}}}{\widetilde C_{R_z}},
\end{equation}
where $\widetilde C_{R_z}$ is the number of arbitrary-angle $R_z$ rotations in one simulated $U(\tau)$.

Before extracting gate counts and depths, barriers are removed and circuits are optimized with the \texttt{pytket} passes
\texttt{DecomposeBoxes}, \texttt{RemoveRedundancies}, \texttt{AutoRebase} to $\{\mathrm{CX},R_z,H\}$, \texttt{RemoveRedundancies} , \texttt{RebaseToCliffordSingles}, and a final \texttt{RemoveRedundancies} pass, in the respective order. The Givens-rotation basis-change subcircuits also apply \texttt{RemoveRedundancies} after numerical angle substitution. This performed local cancellations and set the counting basis to CX, arbitrary-angle $R_z$, and one-qubit Clifford gates, but did not include other optimization.

These estimates should be interpreted as conservative, especially for the final $T$ counts. Tighter double-factorized truncation or alternative compression can reduce $L$ and $\Lambda_L$ \cite{Motta_2021,Lee2021,vonBurg2021}, tighter Trotter analysis or optimized product formulas can reduce $n_{\mathrm{Trot}}$ \cite{Poulin2015TrotterStep,Childs2019TheoryTrotter}. Qubitization and other Hamiltonian-simulation methods can further improve the absolute asymptotic scaling beyond this Trotterized model, and improved deterministic or probabilistic/RUS synthesis can reduce the per-rotation $T$ cost \cite{Selinger2015,ross2016optimal,PaetznickSvore2014}. A less conservative spectral window for choosing $\tau$ could also move the total simulated time closer to the Heisenberg-limited scale \cite{LinTong_2022}. These refinements can change the absolute resource estimates by orders of magnitude, but for large-$N$ the gain ratios between SE-QPE and QPE are determined by the relative controlled-evolution, CSWAP-gadget costs and the runtime halving.

\end{document}